# Impact of dynamical regionalization on precipitation biases and teleconnections over West Africa

Iñigo Gómara[1,2,3*], Elsa Mohino[1], Teresa Losada[1], Marta Domínguez[1,2], Roberto Suárez-Moreno[1,2] and Belén Rodríguez-Fonseca[1,2]

[1] *Dpto. Geofísica y Meteorología, Universidad Complutense de Madrid, Madrid, Spain*

[2] *Instituto de Geociencias (IGEO), UCM, CSIC, Madrid, Spain*

[3] *CEIGRAM, Universidad Politécnica de Madrid, Madrid, Spain*

[*]**Correspondence to:** Iñigo Gómara, Dpto. Geofísica y Meteorología, Universidad Complutense de Madrid, Facultad de CC. Físicas, Ciudad Universitaria s/n. 28040 Madrid, Spain. E-mail: i.gomara@ucm.es

**Abstract**

West African societies are highly dependent on the West African Monsoon (WAM). Thus, a correct representation of the WAM in climate models is of paramount importance. In this article, the ability of 8 CMIP5 historical General Circulation Models (GCMs) and 4 CORDEX-Africa Regional Climate Models (RCMs) to characterize the WAM dynamics and variability is assessed for the period July-August-September 1979-2004. Simulations are compared with observations. Uncertainties in RCM performance and lateral boundary conditions are assessed individually.

Results show that both GCMs and RCMs have trouble to simulate the northward migration of the Intertropical Convergence Zone in boreal summer. The greatest bias improvements are obtained after regionalization of the most inaccurate GCM simulations. To assess WAM variability, a Maximum Covariance Analysis is performed between Sea Surface Temperature and precipitation anomalies in observations, GCM and RCM simulations. The assessed variability patterns are: El Niño-Southern Oscillation (ENSO); the eastern Mediterranean (MED); and the Atlantic Equatorial Mode (EM). Evidence is given that regionalization of the ENSO-WAM teleconnection does not provide any added value. Unlike GCMs, RCMs are unable to precisely represent the ENSO impact on air subsidence over West Africa. Contrastingly, the simulation of the MED-WAM teleconnection is improved after regionalization. Humidity advection and convergence over the Sahel area are better simulated by RCMs. Finally, no robust conclusions can be determined for the EM-WAM teleconnection, which cannot be isolated for the 1979-2004 period. The novel results in this article will help to select the most appropriate RCM simulations to study WAM teleconnections.

## 1. Introduction

The monsoon is the most prominent climate feature of Western Africa during boreal summer, and highly determines socio-economic development over this region (Janicot et al. 2011; Parry et al. 2007). First, because most of the annual total rainfall accumulates during a period of few months (Le Barbé and Lebel 1997; Laurent et al. 1998). Second, because West African societies are highly vulnerable to water availability, and its scarcity/excess can lead to strong agricultural losses, famine and disease spread (Dilley et al. 2005; Cook 2008; Yaka et al. 2008).

The West African Monsoon (WAM) is driven by sea-land contrast of temperature and surface pressure between the Gulf of Guinea and the Sahara Desert. Therefore, during the summer months, North African continental areas are heated up more rapidly than oceanic waters (Fontaine et al. 1998; Rowell 2001). Consequently, the Intertropical Convergence Zone (ITCZ) is pushed northward and humidity is primarily advected towards the Sahel by the monsoon flux. To a lower extent, the Saharan Heat Low also plays a role in this process (SHL; Rodríguez-Fonseca et al. 2011). Additional to its strong seasonal cycle, the WAM also shows remarkable variability at interannual and multi-decadal timescales, in both cases mainly due to global Sea Surface Temperature (SST) anomalies (Folland et al. 1986; Hoerling et al. 2006; Biasutti and Giannini 2006; Mohino et al. 2011a; Rodríguez-Fonseca et al. 2011; 2015).

At interannual timescales, three of the most prominent SST patterns influencing the WAM dynamics are: (i) El Niño-Southern Oscillation (ENSO; Janicot et al. 2001; Rowell 2001; Joly and Voldoire 2009; Mohino et al. 2011b; Diatta and Fink 2014); (ii) the eastern Mediterranean (MED; Rowell 2003; Polo et al. 2008; Fontaine et al. 2010); and (iii) the Atlantic Equatorial Mode (EM; Zebiak 1993; Rowell et al. 1995; Janicot et al. 1998; Losada et al. 2010). According to the available literature, during a positive ENSO event, the warmer waters of the equatorial Pacific trigger a Kelvin wave which, over West Africa, is associated with increased air subsidence and reduced rainfall (Semazzi et al. 1988; Moron and Ward 1998; Janicot et al. 2001; Rowell 2001; Mohino et al. 2011b). The contrary holds for a negative ENSO event. This is one of several proposed mechanisms for the ENSO-WAM teleconnection (Joly and Voldoire 2009). For the Mediterranean, a warming in the eastern basin is known to increase evaporation and enhance moisture advection towards the Sahel (Polo et al. 2008; Fontaine et al. 2010; Gaetani et al. 2010). Such effect, combined with the circulation associated with the SHL, and the further penetration of the monsoonal flow over northern Africa, amplifies moisture convergence and precipitation over the Sahel (Rodríguez-Fonseca et al. 2011). Lastly, for the EM-WAM teleconnection, warmer SSTs over the equatorial Atlantic act to mitigate the pressure gradient between the Gulf of Guinea and Sahara during boreal summer. As a consequence, when an isolated positive EM event is present, precipitation is enhanced over the Gulf of Guinea and reduced over the Sahel (Rowell et al. 1995; Janicot et al. 1998; Vizy and Cook 2001; Losada et al. 2010). The opposite holds for a negative phase EM. However, during the last decades of the 20$^{th}$ century, both ENSO and EM patterns have tended to emerge coincidentally and in counter-phase (Rodríguez-Fonseca et al. 2009). In response to both simultaneous forcings, the characteristic dipolar precipitation pattern associated with an isolated positive (negative) EM pattern has turned into a homogeneous mode of enhanced (reduced) rainfall over West Africa (Mohino et al. 2011c; Losada et al. 2012; Suárez-Moreno and Rodríguez-Fonseca 2015).

Within the framework of climate research, coupled Ocean-Atmosphere General Circulation Models (GCMs) are a useful tool to characterize present/future climate dynamics and variability (IPCC AR5). However, due to the high computational costs, dynamical processes are typically resolved over a coarse horizontal grid of around 100-200 km. Therefore, important dynamical features of the WAM such as Mesoscale Convective Systems (MCSs) or fine scale processes (e.g., soil-atmosphere interactions, land use) are disregarded by GCMs (Cook 2008; Steiner et al. 2009; Domínguez et al. 2010; Hourdin et al. 2010; Sylla et al. 2010). Additionally, the existence of large positive SST biases over the equatorial Atlantic in GCMs leads to a too southward representation of the Atlantic ITCZ branch during boreal summer (Richter and Xie 2008; Li and Xie 2012; Xu et al. 2014). In this context, the Coupled Model Inter-comparison Project phase 5 (CMIP5) initiative has recently coordinated efforts between international scientific communities and end users to improve present and future GCM simulations (Taylor et al. 2012).

The growing demand of policy makers for regionalized projections has led to the foundation of similar initiatives based on Regional Climate Models (RCMs), such as the Coordinated Regional Climate Downscaling Experiment (CORDEX; Giorgi et al. 2009; Jones et al. 2011). In this framework, CORDEX Africa prioritizes efforts on dynamical regionalization of the whole African continent (Nikulin et al. 2012). This initiative has motivated several analyses on precipitation dynamics and variability over different African Sectors: South Africa (Meque and Abiodun 2015; Favre et al. 2016), Eastern Africa (Endris et al. 2016), West Africa (Diallo et al. 2013; Gbobaniyi et al. 2014; Diasso and Abiodun 2015; Akinasola et al. 2015; Adeniyi and Dilau 2016), etc. Whereas most of these have focused on RCM simulations driven by ERA-Interim, a few less have evaluated GCM driven regionalizations. In this context, it is well known that RCMs are able to provide added value (AV) in comparison to GCMs over local areas due to resolved high resolution processes (Giorgi and Mearns 1999; Di Luca et al. 2013). However, it is not so clear whether RCMs can generate AV for large-scale processes. Whereas some authors promote the use of large-scale nudging toward the lateral boundary conditions provided by GCMs (Castro et al. 2005; Laprise et al. 2008), others dissuade against such methodology and adduce that large-scale AV can be generated by RCM internal dynamics whenever the regionalized area is sufficiently large (Mesinger et al. 2002; Veljovic et al. 2010). In any case, high resolution outputs can be meaningful only if the simulated large-scale is realistic or, at least, not worse than in the forcing GCM. Hence, both lateral boundary conditions and RCM performance are pivotal aspects for the correct representation of the large-scale circulation, and therefore teleconnections.

In order to shed light upon this debate, the ability of GCMs and RCMs to represent the West African Monsoon and its most prominent teleconnection patterns is analyzed in this article. This is done through the analysis of CMIP5 historical and CORDEX-Africa simulations, accounting for uncertainties in RCM lateral boundary conditions and performance. For this purpose, a Maximum Covariance Analysis (MCA) between seasonal anomalous SSTs and precipitation is performed through the use of the novel Sea Surface Temperature based Statistical Seasonal Forecast model (S$^4$CAST; Suárez-Moreno and Rodríguez-Fonseca 2015).

The present article is organized as follows. Data and Methodology are provided in sections 2 and 3. Section 4 describes the climatological aspects and most prominent teleconnections of the WAM in observations, GCMs and RCMs. The article concludes with a summary and a discussion of the main results.

## 2. Data

In this study both observations and climate model simulations are utilized.

### 2.1 Observations

Three different gridded datasets are used to analyze monthly precipitation and large-scale dynamics over West Africa:

1. The NCEP/NCAR Global Precipitation Climatology Project (GPCP) version 2.1: This dataset starts in January 1979 and has a horizontal resolution of 2.5 degrees in longitude/latitude worldwide. GPCP version 2.1 is based on multi-satellite measurements and gauge observations (Huffman et al. 2009).

2. The NASA Modern Era-Retrospective Analysis for Research and Application (MERRA): Based on the Goddard Earth Observing System version 5 (GEOS-5), it spans from 1979 to the present and covers the entire globe at a horizontal resolution of 0.5/0.67 degrees in longitude/latitude (Rienecker et al. 2011).

3. The European Centre for Medium-Range Weather Forecasts Interim Re-analysis (ERA-Interim): It prolongs from 1979 to the present and features a worldwide horizontal resolution of 0.75 degrees in longitude/latitude along 60 vertical levels (Dee et al. 2011).

Additionally, sea surface temperature and sea ice data from the Met Office Hadley Centre (HadISST) are utilized. This dataset spans from 1870 to the present on monthly basis and has a horizontal resolution of one degree in longitude/latitude. HadISST is based on the Met Office Marine Data Bank and the Global Telecommunications System, among other data sources (Rayner et al. 2003).

*2.2 Model Simulations*

Historical simulations of 8 different GCMs from the CMIP5 initiative are analyzed. The simulations cover the last century and a half (1861-2005) and are mainly forced by observations of atmospheric gas composition (including greenhouse gases) and land cover (Taylor et al. 2012). The model names and modeling institutions are provided in Table 1.

Additionally, 11 RCM simulations from the CORDEX-Africa experiment are studied. All models cover the period 1951-2005 over the African continent at a horizontal resolution of 0.44 degrees in both longitude and latitude (cf. Fig. S1a; Giorgi et al. 2009; Nikulin et al. 2012). The nested RCM simulations are separated into two different experiment blocks depending on the GCM-RCM configuration. First, a set of 8 different GCM historical simulations driving the same regional model (SMHI-RCA4) is considered. This is done to assess uncertainties in RCM boundary lateral conditions. Second, the same GCM model (MPI-ESM-LR) is set as boundary condition for 4 different regional models (CCLM4-8, CSC-REMO, SMHI-RCA4 and UQAM-CRCM5). Here, the performance of different nested RCMs is analyzed based on the same external forcing. For clarity, the GCM-RCM combinations considered in this article are specified in Table 1. Hereafter we will refer as GCMs, GCMs-RCA4, MPI and MPI-RCMs to the GCM-RCM block experiments described in this section.

## 3. Methods

In order to analyze the mean state and teleconnections of the WAM, the peak precipitation period (July-August-September; JAS) over the West African area [WA; 20ºW-30ºE, 0º-20ºN] is selected. The choice of area follows Sultan and Janicot (2000) and Janicot et al. (2011). Additionally, two smaller regions representative of West Africa North [WA-N; 10ºW-10ºE, 9º-15ºN] and West Africa South [WA-S; 10ºW-10ºE, 5º-9ºN] are considered (cf. boxes in Fig. 1a). In these regions, human socio-economic activity is highly linked to precipitation, which is homogeneous all over their territory (cf. CORDEX protocol; Nikulin et al. 2012; Laprise et al. 2013; Gbobaniyi et al. 2014; Dosio et al. 2015). The period of study covers 26 consecutive JAS seasons, from 1979 to 2004, which is the common time interval for model and observational data. Year 2005 is not included in the analysis due to the unavailability of data for some particular models and variables.

*3.1 Interpolation of the fields*

A bilinear interpolation is applied to all fields of this study (e.g., precipitation, Sea Surface Temperature, wind divergence, etc.), which are regridded to a 2.8º x 2.8º horizontal resolution. The grid corresponds to the GCM of coarser resolution (CanESM2). This is done to allow comparison between fields from different sources (GCMs, RCMs and observations) and to avoid issues derived from increasing data resolution artificially (Wilks 2006). Thus, the remapping of RCM fields returns clearer large-scale spatial patterns at the cost of obscuring fine-scale processes. Nevertheless, the main focus of this paper is more on the large-scale atmospheric circulation and less on local-scale features. To keep consistency among calculations, all SST maps are also downgraded to the resolution of CanESM2.

*3.2 Calculation of Ensembles*

Ensemble means are provided along this article for the GCM, GCM-RCA4 and MPI-RCM block experiments. In each experiment block the output from each model simulation (1 realization per model) is averaged over the rest. Robustness of ensembles is calculated from the number of members of each class depicting the same sign of a given variable (e.g., precipitation anomaly, regression map, etc.).

*3.3 Empirical Orthogonal Functions and Maximum Covariance Analysis*

Empirical Orthogonal Functions (EOF) analysis is a powerful discriminant statistical method that determines the most prominent modes of variability of a given anomalous field (Lorenz 1956). In this article, EOF analysis is performed on SST and precipitation anomaly fields. To evaluate the co-variability of West African rainfall with Tropical Pacific, Mediterranean and Tropical Atlantic SST anomalies, a Maximum Covariance Analysis (MCA) is also utilized (Bretherton et al. 1992; Widmann 2005). The latter is often considered as a generalization of EOFs. MCA is performed between a predictor Y field (SST anomalies) and a standardized predictand Z field (precipitation anomalies) over specific regions and time periods. The method applies a Singular Value Decomposition to the non-square covariance matrix of both fields, and performs linear combinations between the time series of Y and Z (a.k.a. expansion coefficients U and V, respectively) to maximize the former (Cherry 1997). Once the covariance matrix is diagonalized, the method retrieves the singular vectors R and Q, which are the leading co-variability modes between Y and Z over the regions and period selected. The squared covariance fraction (scf) of these modes is provided in %, with a confidence interval (90%) calculated through a Monte Carlo test of 1000 random iterations. The linear trend is removed from the SST predictor/predictand fields to mitigate the potential signal of anthropogenic climate change in the MCA analysis (IPCC AR5). No further filtering is applied to the fields, as the aim is to focus on interannual variability, and potential lower-frequency oscillations (e.g., decadal, multi-decadal) are hard to detect over a period of just 26 years. For the same reason, it is also assumed that teleconnections are stationary during the study period, despite it does not seem to be the case when longer intervals are considered over specific areas: e.g., West Africa (Mohino et al. 2011c), the Tropical Atlantic (Rodríguez-Fonseca et al. 2009; 2016), the extra-tropical North Atlantic (López-Parages and Rodríguez-Fonseca 2012; Raible et al. 2014; Gómara et al. 2016), etc.

For the calculation of the MCA, the Sea Surface Temperature based Statistical Seasonal Forecast model (S$^4$CAST; v.2.0), recently developed by Suárez-Moreno and Rodríguez-Fonseca (2015), is used. Although the S$^4$CAST model was mainly conceived to work in 'forecast' mode (predictor field leading in time), its 'synchronous' mode is selected (lead time 0), as the objective is to accurately characterize concurrent climate anomalies in observations, GCMs and RCMs.

*3.4 RCM Added Value*

RCM Added Value (AV) is calculated throughout this article for different variables. This measure provides information on how well RCMs are able to reproduce a given field in comparison with GCMs, considering gridded observations as basis. The AV is computed as follows (Di Luca et al. 2013; Meque and Abiodun 2015):

$$AV = (X_{GCM} - X_{OBS})^2 - (X_{RCM} - X_{OBS})^2, \quad (1)$$

Where $X_{GCM}$, $X_{RCM}$ and $X_{OBS}$ are the corresponding fields for the GCM, RCM and observations, respectively. Thus, if AV is positive (negative), the RCM improves (deteriorates) GCM simulations, taking observations as reference.

**4. Results**

*4.1 Seasonal Precipitation Biases*

In this section, seasonal precipitation biases of GCM, GCM-RCA4, and MPI-RCM simulations are analyzed for the period JAS 1979-2004. The seasonal mean precipitation of the observational GPCP dataset, which is chosen as reference, is provided in Fig. 1a. The characteristic zonal rainfall belt extending from West Africa to Chad/Sudan is present in the figure, with two regional maxima situated over the Guinea-Conakry/Sierra Leone coast and the Cameroon Highlands (Cook and Vizy 2006). The standard deviation field appears overlaid in Fig. 1a. The spatial overlapping between the mean and standard deviation indicates that the strongest variability is enclosed over the areas of highest precipitation. These patterns are similar to that obtained from MERRA (Fig.

S1b). Nevertheless, it is worth to mention that GPCP tends to produce higher precipitation rates over the Central African Republic, Chad and surrounding area (Fig. S1c). This inconsistency has been mentioned in previous studies and appears to be caused by different gauge station availability over the area in GPCP and MERRA datasets (Huffman et al. 2009; Yin and Gruber 2010; Nikulin et al. 2012).

Next, the ensemble precipitation bias of GCM historical runs is shown (Fig. 1b). As it can be observed, GCMs have difficulties to accurately represent the northward migration of the ITCZ in boreal summer and its associated rain belt. As a consequence, rainfall amounts and variability are notably overestimated south of the Gulf of Guinea coast. The contrary is observed over the westernmost Sahel, especially over Senegal and Gambia. As mentioned in the introduction, this is caused by the existence of warm SST biases over the tropical Atlantic in GCMs (Fig. S1d), which act to mitigate the sea-land pressure gradient between the Gulf of Guinea and the Sahara Desert (Richter et al. 2012). For completeness, the individual seasonal precipitation biases of GCM historical runs are presented in Figs. S2a-h. In this context, HadGEM2-ES (Fig. S2e) and NorESM1-M (Fig. S2h) are clear examples of a too southward representation of the ITCZ. Oppositely, MPI-ESM-LR (a.k.a. MPI) provides one of the best estimates of the WAM over West Africa (Fig. S2g).

For comparison, the ensemble precipitation bias of GCM-RCA4 simulations is provided in Fig. 1c. Rainfall is again overestimated over the Gulf of Guinea and underestimated over Senegal. However, the zonal band of inflated precipitation appears stronger and narrower (compare Figs. 1b and 1c). In addition, precipitation is evidently underestimated over the Congo region (Fig. 1c), an aspect not observed in GCMs (Fig. 1b). Such error is present systematically in all GCM-RCA4 individual simulations (Figs. S2i-p), and may be inherent to the SMHI-RCA performance. Although for a slightly different time period, model version and lateral boundary conditions (ERA-Interim), the same dry bias over the Congo area was observed between SMHI-RCA35 and GPCP in Nikulin et al. (2012; their Fig. 4).

Next, the ensemble of MPI-RCMs seasonal precipitation bias and its individual components are provided in Figs. 1d and S2q-t, respectively. The selection of MPI as boundary condition for the RCM runs is based on the relative good performance of this model representing the WAM (Fig. S2g). Consistent with the previous GCM-RCA4 simulations, most of RCMs tend to exaggerate precipitation south of the Guinean coast (Fig. 1d). Regarding the dry bias over the Congo region, it is again evident in the MPI-RCM ensemble, although of weaker intensity (compare Figs. 1c and 1d). Attending to the individual runs (Figs. S2q-t), this appears to be a common issue in RCMs, and not only inherent to SMHI-RCA4.

Finally, the RCM added value of seasonal precipitation is provided in Figs. 1e-f (ensembles) and Fig. S3 (individual runs). For the GCMs-RCA4 ensemble, the regional model is able to minimize mean precipitation biases inland over West Africa and the Tropical South Atlantic (Fig. 1e). Average negative AVs are present south of the Guinean coast and over the Congo area. Individual AVs of GCM-RCA4 runs are available in Figs. S3a-h. Both HadGEM2-ES/RCA4 (Fig. S3e) and MIROC5/RCA4 (Fig. S3f) runs are examples of good performance over West Africa and the Gulf of Guinea. The opposite case is GFDL-ESM2M/RCA4 (Fig. S3d), which remarkably fails to represent precipitation south of the Guinean coast and is the main contributor of the ensemble negative AVs over this area in Fig. 1e. Regarding MPI-RCM simulations (Fig. 1f), the ensemble AV shows negative values all over West Africa's coast and mainland. The scarce positive values are present over the equatorial and Tropical South Atlantic. Attending to the equation of Added Value exclusively (1), one possibility for these results is that the MPI simulation already provides quite accurate values of the WAM rainfall amounts (Fig. S4g). Consequently, little room is left for RCM improvement (Meque and Abiodun 2015). Nevertheless, the key factor behind GCM/RCM precipitation biases is the role of physics in the simulations considered. Due to resolution limitations, convective rainfall is parameterized in both global and CORDEX-Africa simulations analyzed, but the convection scheme may not be similar between the driving GCM and the regional model in some cases (Nikulin et al. 2012). As rainfall variability is highly sensitive to the physical package utilized for convection (and other parameterized processes; Flaounas et al. 2011), a significant/systematic improvement of precipitation biases in all RCM simulations is hard to be accomplished. Anyhow, these conclusions may be different in convection-permitting or cloud-resolving higher resolution simulations, where a significant improvement should be expected after regionalization (Randall et al. 2003; Prein et al. 2015). Lastly, the individual AVs of MPI-RCM simulations are available in Figs. S3i-l.

So far, precipitation biases have been spatially characterized over West Africa. Subsequently, average precipitation biases over West Africa North [WA-N; 10ºW-10ºE, 9º-15ºN], West Africa South [WA-S; 10ºW-10ºE, 5º-9ºN] and the whole West Africa domain [WA; 20ºW-30ºE, 0ºN-20ºN; cf. Sections 4.2 to 4.4] are quantified in Fig. 2. The mean values corresponding to the MERRA dataset in both regions are included as well in Fig. 2. In general, GCM-RCA4 simulations tend to reduce mean seasonal precipitation biases of GCMs over WA-N, WA-S and WA. These results can be inferred from the 'Ensemble GCMs' bars in Figs. 2a,c,e and are valid for both GPCP and MERRA. The contrary holds for MPI and MPI-RCMs over WA-S and WA, where precipitation biases are increased after regionalization. This can be extracted from the 'Ensemble MPI' bars of the same figures (GPCP and MERRA). For WA-N, the mean seasonal precipitation biases are fairly weak in both MPI and MPI-RCM simulations. In general terms, results for the whole WA are similar to those from WA-S, but with lower values (Figs. 2c,e).

Regarding standard deviation, most of GCM-RCA4 members tend to underestimate the year-to-year variability in seasonal precipitation over WA-N (Fig. 2b), while the reverse holds for WA-S (Fig. 2d). Results are disparate for the whole WA domain (Fig. 2f). Regarding the MPI-RCMs, most of members overestimate the year-to-year variability in seasonal precipitation in WA-S and WA (Figs. 2d,f). Among them, MPI-CCLM4 and MPI-REMO biases largely exceed the observational uncertainty in all regions, suggesting that these two models control the precipitation variability of the 'MPI Ensembles'. Due to the large differences in standard deviation observed between GPCP and MERRA over all regions (horizontal red lines; Fig. 2 - right column), no further conclusions can be here obtained.

So far, climatological GCM and RCM simulations of the WAM have been compared with observations. In the following, a similar analysis is carried out but focusing on the representation of the most prominent interannual variability SST patterns influencing the WAM (cf. Section 1).

*4.2 Influence of El Niño-Southern Oscillation on the West African Monsoon in observations, GCMs and RCMs*

In this section the impact of ENSO on interannual WAM variability is assessed in observations, GCMs and RCMs. For this purpose, a MCA between seasonal SST anomalies over the Equatorial Pacific [110°E-80°W, 20°S-20°N; predictor field] and simultaneous precipitation anomalies over West Africa [20°W-30°E, 0°-20°N; predictand field] is performed. The linear trend is removed from the predictor/predictand fields to mitigate the potential influence of anthropogenic climate change in the results.

*(A) Observations*

In Figs. 3a-b the SST homogeneous and precipitation heterogeneous maps are provided for the HadISST-GPCP gridded observational datasets. The leading SST mode depicts a well-defined and robust positive phase ENSO pattern (shadings/stippling in Fig. 3a), which accounts for 41% of total squared covariance fraction (scf) and 58% of explained SST variance over the Equatorial Pacific (EOF1; cf. Fig. S4a and Table S1). The pattern in Fig. 3a is accompanied in the Equatorial Atlantic by much weaker negative SST anomalies (Rodríguez-Fonseca et al. 2009; Losada et al. 2012). The associated heterogeneous map reveals a large-scale and statistically significant pattern of decreased precipitation that extends along the corridor 0°-20°N over Western and Central Africa (Fig. 3b). This pattern corresponds to EOF1 of seasonal precipitation over West Africa (precipitation exp. var. 30%; cf. Fig. S4d and Table S2). By construction, the co-variability modes are also valid if a minus sign is applied to both maps (i.e., negative ENSO phase & increased precipitation over Western/Central Africa). In order to characterize the underlying dynamics of the ENSO-WAM teleconnection, the projection of expansion coefficient U on two different dynamical fields from ERA-Interim is given in Fig. 3c. The selected fields are anomalous seasonal wind divergence (shadings) and velocity potential (contours). To characterize the fields at both lower and upper levels, anomalies from pressure level 850 hPa are subtracted to those from 200 hPa (hereafter DIV200/850 and KHI200/850). Therefore, positive regression anomalies of DIV200/850 will generally be associated with enhanced air divergence at 200 hPa and convergence at 850 (i.e., increased air uplift). The contrary holds for negative DIV200/850 anomalies (enhanced air subsidence). For velocity potential, positive (negative) anomalies are associated with intensified downward (upward) air movement.

Attending to Fig. 3c, a positive ENSO event is linked to enhanced air subsidence over the Gulf of Guinea and the West Africa corridor 0°-20°N. Please note that the color bar in Fig. 3c is reversed to improve visual comparison with Fig. 3b. To permit a wider perspective, the regression map of KHI200/850 is provided globally in Fig. 3a (contours). As expected, the outcome from Figs. 3a-c is in good agreement with the so-called "ENSO-WAM Kelvin wave teleconnection mechanism", already described in previous studies (Folland et al. 1986; Palmer 1986; Janicot et al. 1996; 2001; Rowell 2001; Joly and Voldoire 2009; Mohino et al. 2011b). Finally, the negative SST anomalies present over the equatorial Atlantic might, at some point, have an imprint on West Africa precipitation. However, the ENSO related SST anomalies are of much stronger intensity and seem to dominate the teleconnection (Fig. 3). This is consistent with Losada et al. (2012).

*(B) GCMs & GCMs-RCA4*

As a next step, the ENSO-WAM teleconnection is assessed on the historical GCM runs (Joly et al. 2007). For simplicity, ensemble GCM maps are discussed first and provided in the main manuscript (Fig. 4, left column). Results for individual simulations are available in the supplementary material (Fig. S5).

The GCM ensemble homogeneous SST and heterogeneous precipitation maps are provided in Figs. 4a,c. On the one hand, a robust positive ENSO pattern (32% scf), very similar to that obtained from observations, is present in Fig. 4a (stippling indicates 7/8 models with same sign on regression). A concomitant negative phase EM over the eastern tropical Atlantic is also present. On the other hand, two areas of decreased precipitation can be seen in the ensemble heterogeneous map (Fig. 4c). The first is situated over the Equatorial Atlantic. The second stretches along 15°N over Western Africa. The projection of expansion coefficient U on anomalous DIV200/850 and KHI200/850 fields is provided for the GCM ensemble (Fig. 4e). The spatial overlapping between precipitation and DIV200/850 anomalies in Figs. 4c,e is very clear and corroborates the strong relation between both fields. Therefore, GCMs are also able to reproduce the ENSO-WAM teleconnection mechanism detected in observations.

However, the strongest precipitation/subsidence anomalies are shifted to the south in GCMs (compare Figs. 3b,c and 4c,e). That is consistent with a too southward representation of the ITCZ (Fig. 1b). In addition, the enhanced subsidence and decreased precipitation anomalies over the Sahel are narrower and of weaker intensity in GCMs. Regarding the individual GCM simulations (Fig. S5), in all of them ENSO represents the leading mode, with squared covariance fractions ranging from 25% (CNRM-CM5) to 45% (MIROC5). For the individual precipitation and subsidence maps, a certain degree of dissimilarity among models is found. For instance, CNRM-CM5 is associated with increased precipitation and air ascent over the Guinean coast during positive ENSO events (Figs. S5d-f), whereas the anomalies are of opposite sign in HadGEM2-ES and MIROC5 under a similar external forcing (Figs. S5m-r).

After having characterized the ENSO-WAM teleconnection in observations (Fig. 3) and GCMs (Fig. 4-left column), the same analysis follows for the ensemble of GCM-RCA4 simulations (Fig. 4-right column). On the one hand, the SST ensemble homogeneous map from GCMs-RCA4 is exceptionally similar to that found from GCMs (cf. Figs. 4a and 4b). That is expected as the MCAs for the GCM-RCA4 simulations are based on the same GCM predictor data (only precipitation is regionalized over Africa) and the dominance of ENSO variability over the equatorial Pacific is paramount (Trenberth 1997). On the other hand, the precipitation ensemble of GCM-RCA4 heterogeneous maps is not robust in space (Fig. 4d), and only depicts multi-model agreement near the equator. Attending to the regression maps of DIV200/850 and KHI200/850, the outcome is very similar (Fig. 4f). The ENSO impact on air subsidence only remains robust near the equator and the signal over the Sahel vanishes. In this line, a strong multi-model spread is found in individual precipitation and subsidence GCM-RCA4 regression maps (Fig. S6). For instance, results from CanESM2/RCA4 (Figs. S6b,c) and GFDL/RCA4 (Fig. S6k,l) are very different over the Sahel. The squared covariance fraction of the leading MCA modes from GCM-RCA4 simulations also returns more spread in their values (21 to 54%).

*(C) MPI & MPI-RCMs*

Next, the ENSO-WAM teleconnection is assessed for MPI and MPI-RCM simulations. The analysis is the same as provided in (B), with the exception that MPI results (Fig. 5, left column) are specific to a single model simulation, whereas MPI-RCM ensembles are composed of 4 members (Fig. 5, right column).

Attending to MPI performance, it relates an ENSO type of SST anomaly (Fig. 5a; scf 33%) with decreased precipitation (Fig. 5c) and increased air subsidence (Fig. 5e; cf. DIV200/850 in colors) over the Gulf of Guinea. However, the simulated ENSO impact over the Sahel (15ºN) is very weak, especially near the Greenwich meridian (Figs. 5c,e).

Regarding MPI-RCM results (Fig. 5, right column), the ensemble SST regression map (Fig. 5b; scf 31%) is almost identical to that obtained from MPI (Fig. 5a). This is because all simulations use the same SST forcing from MPI (cf. individual SST maps in Fig. S7). For the precipitation and air subsidence ensemble maps (Figs. 5d,f), results are very similar to those from the MPI simulation. In particular, the ENSO impact on Sahelian precipitation is even weaker in the MPI-RCMs ensemble (compare Figs. 5c and 5d). In this line, MPI-RCM individual simulations return rather different precipitation/subsidence anomalies over Western Africa (Fig. S7), as are the cases of MPI/CCLM4-8 (Figs. S7a-c) and MPI/CSC-REMO (Figs. S7d-f).

*(D) Added Value results*

As a summary of sections (A)-(C), the ensemble RCM added value of heterogeneous precipitation maps is calculated based on Eq. (1). For this purpose, the average AV from individual simulations (shadings) and the number of models in which this variable is positive at a given grid-point (contours) are shown in Fig. 6. In this case, positive RCM AVs reveal regions where the representation of the ENSO-WAM teleconnection is improved compared with GCMs (considering observations as basis).

The results of GCM-RCA4 vs. GCM simulations are provided in Fig. 6a. AVs are mainly negative over the box of study [20ºW-30ºE, 0º-20ºN], and the scarce positive values appear constrained over the Equator. Very similar results are obtained for MPI-RCMs vs. MPI (Fig. 6b). Therefore, RCMs seem to deteriorate the representation of the ENSO-WAM teleconnection provided by GCMs (see also individual AV members in Fig. S8). In this context, GCMs are able to capture, although with weaker intensity, the large-scale ENSO impact on air subsidence over the Sahel (Figs. 3-4). However, RCMs are not able to improve the signal present in the lateral boundary conditions over their domain (Figs. 4-5). According to these results, the considered CORDEX-Africa RCMs are less skillful than GCMs to represent the ENSO-WAM teleconnection. At this point, it must be reminded that this study does not intend to clarify the technical reasons why RCMs improve/deteriorate GCM simulations. The main focus is to evaluate how well interannual variability modes are simulated by GCMs/RCMs and under which circumstances these simulations could be used in future research.

*4.3 Influence of the Mediterranean on the West African Monsoon in observations, GCMs and RCMs*

This section focuses on the evaluation of the MED-WAM teleconnection in observations, GCMs and RCMs. JAS seasonal SST anomalies over the whole Mediterranean domain [0º-40ºE, 30ºN-45ºN] are considered as predictor. Precipitation anomalies over the West African domain [20ºW-30ºE, 0º-20ºN] are selected as predictand.

*(A) Observations*

Fig. 7 depicts the MCA results between HadISST and GPCP observational datasets. The leading co-variability mode explains 40% of squared covariance fraction. The homogeneous SST map shows a robust SST warming in the Mediterranean (Fig. 7a). This pattern corresponds to EOF1 of SST anomalies over the same area and explains 52% of variability (Fig. S4b and Table S1). The heterogeneous rainfall map exhibits widespread, significant positive anomalies over the Sahel (Fig. 7b). Expansion coefficient V of this pattern is significantly linked to EOF1 (exp. var. 30%; Fig. S4d) and EOF2 (exp. var. 25%; Fig. S4e) of precipitation anomalies over West Africa (Table S2). By linearity of the method, the opposite MCA patterns can also be considered. The dynamics of the WAM-MED teleconnection are characterized in Fig. 7c, where expansion coefficient U is projected on specific humidity at 850 hPa (SHUM850, shadings), moisture flux at 850 hPa (MF850, arrows), and sea level pressure (SLP, contours) anomalies. Specifically, a warming in the Mediterranean is associated with local increased evaporation and low-level moisture, which is advected to the south by the low pressure anomalies situated over the eastern Sahara (Fig. 7c). Concurrently, the anomalous large-scale configuration strengthens the SLP gradient between the Gulf of Guinea and Sahara, thus intensifying the southwesterly monsoonal flow. In consequence, moisture supply and rainfall are increased over the Sahel area (Figs. 7b-c). These results are consistent with previous studies on this topic (Rowell 2003; Jung et al. 2006; Fontaine et al. 2010; 2011; Gaetani el al. 2010).

*(B) GCMs & GCMs-RCA4*

The reproduction of the leading SST mode by the ensemble of individual GCMs is presented in Fig. 8a. A robust warming over the Mediterranean is found (32% scf). A positive warming over the Equatorial Atlantic is also visible in the figure. The latter could be related to the ensemble rainfall positive anomalies present over the Gulf of Guinea region in Fig. 8c. In GCMs, the center of the associated low-pressure anomalies is shifted northward (over the Mediterranean Sea) compared to observations (Figs. 8e and 7c). As a consequence, the anomalous large-scale configuration no longer promotes moisture inflow over the Sahel from the Mediterranean (Fig. 8e). Additionally, the anomalous southwesterly monsoonal flow is clearly underestimated in GCM simulations, probably due to the appearance of EM+ SST anomalies in the ensemble homogeneous map (Fig. 8a). As a result, humidity supply and precipitation are notably reduced over the Sahel area compared to observations (cf. Figs. 7b,c and Figs. 8c,e). On individual GCM simulations (Fig. S9), SST maps generally capture a warming in the Mediterranean. Contrastingly, the positive signal over the tropical Atlantic is not reproduced by all models. An exception is the NorESM1-M model, which provides very similar results to that found in observations (Figs. S9v-x). In this model, the prominent low pressure anomalies over northern Africa enhance low-level moisture advection from the western Mediterranean and Gulf of Guinea towards the Sahel (Fig. S9x). In this line, a correct representation of the Saharan heat low seems essential to characterize the EM-WAM teleconnection (Lavaysse et al. 2010ab; Evan et al. 2015).

Results for GCM-RCA4 simulations are given in Fig. 8 (right column). On the one hand, the ensemble SST homogeneous map reveals again MED+ and EM+ concurrent anomalies (Fig. 8b). On the other hand, the associated precipitation map returns widespread rainfall anomalies over the Sahel area, additional to those over the Gulf of Guinea (Fig. 8d). In this case, southwesterly moisture advection over the Sahel is better simulated than in GCMs, consistent with a strengthened Saharan/Mediterranean heat low (compare Figs. 8e-f). However, the position of the low is again improperly simulated in GCMs-RCA4, and the Mediterranean branch of moisture advection over the Sahel is absent. Thus, precipitation and SHUM850 anomalies are better simulated over the eastern Sahel, where the monsoonal moisture branch appears to dominate (Figs. 8d,f). For completeness, GCM-RCA4 individual regression maps are provided in Fig. S10. The models that better capture the MED-WAM teleconnection are CNRM-CM5/RCA4 (Figs. S10d-f) and NorESM1-M/RCA4 (Figs. S10v-x). In both cases, a robust SST warming is found over the Mediterranean, while anomalies are inexistent over the Equatorial Atlantic. Moreover, in these two models the associated SLP pattern shows a strong latitudinal gradient over the Sahel, a factor that seems crucial to foster inflow humidity flux from the Gulf of Guinea.

*(C) MPI & MPI-RCMs*

The results for MPI and MPI-RCM simulations are collected in Fig. 9. The analysis is conducted as in section (B). In both experiments, ensemble homogeneous maps depict a warming of Mediterranean SSTs (Figs. 9a-b). Compared to GCMs, the concurrent positive anomalies over the equatorial Atlantic are much weaker and confined near the coast (cf. Figs. 8a-b and 9a-b). Attending to the ensemble precipitation regression maps, regionalized simulations are again able to provide a better estimate of precipitation anomalies over the Sahel (Figs. 9c-d). Whereas in the MPI simulation a precipitation dipole is found between the Guinean coast and Sahara (Fig. 9c), in the MPI-RCM ensemble widespread positive anomalies are present over the Sahel (Fig. 9d). In both cases, the limited positive precipitation anomalies over the eastern Equatorial Atlantic could be related to the SST warming over this area. The dynamics of the MED-WAM teleconnection are analyzed in Figs. 9e-f. On MPI (Fig. 9e), the anomalous large-scale SLP configuration seems to promote shallow moisture transport from the Mediterranean towards the Sahel. However, moisture advection from the Gulf of Guinea is completely blocked. As a consequence, neither low-level humidity nor precipitation appear enhanced over the Sahel. On MPI-RCMs (Fig. 9f), the ensemble regression map reveals that moisture advection over the Sahel takes places throughout the northern (Mediterranean) and southern (monsoonal) branches. As a result, Sahelian low-level humidity is significantly enhanced and precipitation is promoted. In this case, the deeper/broader negative SLP anomalies simulated by RCMs over Northern Africa might contribute to strengthen the monsoonal flow (compare Figs. 9e and 9f). According to these results, the southern branch of humidity advection seems to play a much more determinant role than the northern one on the MED-WAM teleconnection.

Regarding individual MPI-RCM simulations (Fig. S11), the warming of the Mediterranean is well captured by the 4 members of the ensemble. Some differences appear in other regions though. For instance, compare MPI/CCLM4 and MPI/CRCM5 runs over the sub-tropical North Atlantic (Figs. S11a and S11j). The less realistic simulation of the MED-WAM teleconnection is given by MPI/REMO (Figs. S11d-f), which completely fails to simulate the zonal SLP gradient over West Africa.

*(D) Added Value results*

The use of nested RCMs in reproducing the MED-WAM teleconnection is weighted by the ensemble added values of heterogeneous rainfall maps (Fig. 10). As expected, robust positive AVs are provided by GCM-RCA4 and MPI-RCM ensembles over the central and eastern Sahel. Over these areas, the Mediterranean influence on precipitation is known to be higher (cf. Rowell 2003; Fontaine et al. 2010; Gaetani el al. 2010). For completeness, AVs for individual members can be examined in Fig. S12. Based on these results, RCMs can be considered as a useful tool to substantially improve GCM simulations of the MED-WAM teleconnection.

*4.4 Influence of the Atlantic Equatorial Mode on the West African Monsoon in observations, GCMs and RCMs*

In this section the Atlantic EM influence on WAM precipitation is analyzed. With this aim, seasonal SST anomalies over the equatorial Atlantic [60ºW-20ºE, 20ºS-5ºN] are selected as predictor field. For the predictand, the same area [20ºW-30ºE, 0º-20ºN] of standardized concurrent precipitation anomalies is chosen.

*(A) Observations*

The MCA results for the HadISST-GPCP observational datasets are shown in Fig. 11. The SST homogeneous map reveals that the Atlantic EM (positive phase) is the leading co-variability mode, explaining 41% of squared covariance fraction (Fig. 11a). This pattern corresponds to EOF1 of SST variability over the Equatorial Atlantic (SST exp. var. 64%; Fig. S4c and Table S1). Together with a positive EM, a negative ENSO pattern is also visible in Fig. 11a over the western equatorial Pacific (Rodríguez-Fonseca et al. 2009). Unlike the SST forcing in Section 4.2 (Fig. 3a), the SST anomalies over the equatorial Atlantic and Pacific are now of similar amplitude.

The associated precipitation regression map is given in Fig. 11b, and depicts a homogeneous mode of positive anomalies extending from the Gulf of Guinea towards the Western Sahel (Losada et al. 2012). Expansion coefficient V of this pattern is significantly linked to EOF1 (exp. var. 30%; Fig. S4d) and EOF2 (exp. var. 25%;

Fig. S4e) of precipitation anomalies over West Africa (cf. Table S2). To analyze the underlying dynamics of this mode, seasonal anomalies of DIV200/850 (colors), MF850 (arrows) and SHUM850 (contours) are regressed on expansion coefficient U (Fig. 11c). As it can be observed, the warmer waters of the equatorial Atlantic (Fig. 11a) increase low-level specific humidity over this area (contours in Fig. 11c). As the pressure gradient between the Gulf of Guinea and Sahara is mitigated due to the Atlantic equatorial SST warming, the northward advection of moisture is restricted over the Guinean coast and surrounding area (arrows in the same figure; Janicot 1992; Fontaine and Janicot 1996). As a consequence, both DIV200/850 (Fig. 11c - colors) and precipitation (Fig. 11b) appear enhanced over this region. In this line, the presence of significant concurrent negative ENSO anomalies may also contribute to increase precipitation and air uplift over the Gulf of Guinea coast (see Fig. 11a, where KHI200/850 regression anomalies are shown in contours). Over the Sahel, the underlying dynamics of the leading precipitation pattern in Fig. 11b are harder to interpret. On the one side, a positive EM forcing tends to produce dryer conditions over this region (Losada et al. 2010; 2012). On the other side, an ENSO- forcing is known to increase air uplift and precipitation over West Africa (cf. Fig. 3). Thus, Sahelian results in Figs. 11b-c seem to be a blend of the aforementioned dynamical mechanisms. This is supported by the strong anticorrelation value (-0.38; 90% confidence interval) obtained between the Atlantic (EM-WAM; Section 4.4) and Pacific (ENSO-WAM; Section 4.2) expansion coefficients U (cf. Table S3).

*(B) GCMs & GCMs-RCA4*

The ensemble map of SST leading patterns obtained from individual GCMs is shown in Fig. 12a. The figure shows a robust SST warming constrained over the eastern equatorial Atlantic (scf 36%), and a well-defined ENSO- pattern over the Pacific. The outcome for individual models is provided in Fig. S13. In general, individual SST maps tend to agree in a warming pattern over the equatorial Atlantic together with statistically significant colder SSTs in the Pacific (5 out of 8 cases; cf. Table S3). However, the area covered by the Atlantic SST anomalies differs notably among models (compare CNRM-CM5 and EC-EARTH; Figs. S13d and S13g). Such limitations may be due to the systematic SST errors in the tropical Atlantic simulated by GCMs, which hamper the representation of EM variability (Fig. S1d; Richter and Xie 2008; Xu et al. 2014).

Regarding the precipitation heterogeneous map, the model ensemble reveals a quite robust pattern of increased precipitation over the eastern equatorial Atlantic, and decreased over Gambia/Senegal (Fig. 12c). Specifically, this precipitation pattern is exceptionally similar to the one obtained in Fig. 1b (GCM ensemble bias). This is not accidental, as in both figures a SST warming in the equatorial Atlantic (caused by either natural variability or GCM errors) leads to a southward shift of the ITCZ. The precipitation pattern in Fig. 12c is also highly consistent with Fig. 12e, where DIV200/850, MF850 and SHUM850 ensemble anomalies are regressed on expansion coefficient U. Over the Gulf of Guinea (Fig. 12e), the combined effect of EM+ (via enhanced moisture supply and air uplift) and ENSO- (enhanced air uplift) is consistent with positive precipitation anomalies (Fig. 12c). These anomalies are, however, shifted to the south in GCMs compared to observations due GCM SST biases. Over the Sahel, ensemble GCM precipitation is clearly reduced compared to observations (cf. Figs. 12c and 11b). One possibility to this behavior is that the influence of ENSO- is not strong enough over the Sahel to counteract the negative precipitation anomalies associated with a positive EM (Losada et al. 2010). This hypothesis is in line with our results in Section 4.2, where the ENSO influence on Sahelian precipitation is shown to be weaker in GCMs than in observations. To complement GCM ensembles, individual regression maps of precipitation and additional fields (KHI200/850, DIV200/850, MF850 etc.) are provided in Fig. S13. Overall, results are diverse among GCM simulations, being CNRM-CM5 a clear example of unrealistic modeling (Fig. S13e-f).

Subsequently, GCM-RCA4 results are provided in Fig. 12 (right column). The ensemble SST homogeneous map returns a very similar pattern to that obtained from GCMs alone (compare Figs. 12a-b). Again, concurrent EM+ and ENSO- SST anomalies are present in the individual regression maps (Fig. S14). Apart from CNRM-CM5/RCA4 (Fig. S14d), the Atlantic EM SST variability is reasonably captured by the individual GCM-RCA4 simulations. In Fig. 12d the ensemble of GCM-RCA4 precipitation maps is provided. The pattern looks even more dipolar than the obtained for GCMs alone, with stronger/broader negative precipitation anomalies over the Sahel (compare Figs. 12c-d). According to Figs. 12e-f, the Sahelian dryer conditions in GCMs-RCA4 are linked to increased air subsidence (compare DIV200/850 anomalies in colors). These results also agree with our findings in section 4.2, where it is shown that the ENSO influence on Sahel precipitation via air subsidence is absent in GCMs-RCA4 (cf. Fig. 4). Therefore, ensemble regression patterns in Figs. 12d,f depict a more dipolar structure, typically associated with an isolated positive EM forcing (Losada et al. 2010; 2012). For completeness, regression maps of individual GCM-RCA4 simulations are also available in Fig. S14.

*(C) MPI & MPI-RCMs*

Next, the EM-WAM teleconnection is assessed on MPI (Fig. 13 - left column) and the ensemble of MPI-RCM runs (Fig. 13 - right column). The homogeneous SST map for MPI (Fig. 13a) reveals that the leading SST mode is EM+ (scf 32%). Although EM+ is accompanied by simultaneous negative SST anomalies over the Tropical Pacific, the latter are very weak (colors) and do not appear to influence the large-circulation over West Africa (KHI200/850 - contours). As a consequence, the associated precipitation heterogeneous map is markedly dipolar, with negative rainfall anomalies over the Sahel and positive in the Gulf of Guinea (Fig. 13c). Regression anomalies of DIV200/850, MF850 and SHUM850 on U are also consistent with a nearly isolated forcing of EM+ (Fig. 13e). First, because enhanced humidity and vertical motion of air are present over the Gulf of Guinea. Second, because stronger air subsidence and dryer conditions are situated over the Sahel. These anomalies are consistent with inhibited moisture flux from the equatorial Atlantic towards continental northern Africa (cf. arrows in Fig. 13e). Subsequently, the outcome of MPI-RCM simulations is provided in Fig. 13 (right column). As expected, the ensemble homogeneous SST map is almost identical to that obtained from MPI alone (compare Figs. 13a-b). Regarding the individual simulations, in 3 out of 4 cases the SST anomalies over the central Pacific are not statistically significant (Fig. S15). As a consequence, ensemble regression anomalies in Figs. 13d (precipitation) and 13f (DIV200/850, SHUM850) are also markedly dipolar between the Gulf of Guinea and Sahel. Nevertheless, for the MPI-RCM ensemble, the anomalies appear attenuated over the Sahel area (compare Figs. 13c,e and 13d,f). The explanation to this behavior is not trivial, especially with the methodology used in this article. Thus, it can only be cautiously conjectured. On the one hand, the MPI/CCLM4 simulation seems to capture some statistically significant ENSO- anomalies accompanying EM+ in the MCA analysis (Fig. S15a). In this simulation, both regressed precipitation (Fig. S15b) and DIV200/850 (Fig. S15c) appear enhanced over the western Sahel, an aspect not observed in the rest of MPI-RCM simulations. Therefore, CCLM4 seems able, for some unknown reason, to amplify the ENSO influence on the WAM compared with the driving GCM model (MPI; Figs. 13c,e). On the other hand, dryer conditions over the Sahel area are found in the regression map of MPI (Fig. 13e - contours) compared with the MPI-RCM ensemble (Fig. 13f). Thus, a better representation of fine-scale processes of the Sahel area (e.g., soil-atmosphere interactions, land use etc.), which may be absent in the MPI run (Steiner et al. 2009; Domínguez et al. 2010; Paeth et al. 2011), could also help to improve simulations of SLP/moisture gradients/precipitation and lead to these results.

To confirm the hypothesis above, several additional sensitivity experiments should be carried out using different GCM/RCM simulations. In particular, it would be interesting to calculate the MCAs setting as predictor an isolated EM+ SST pattern. However, such analysis is far beyond the scope of this paper and might be limited by the occurrence of isolated EM+ anomalies in observations (our study period spans only 26 years). Hence, it is left out for future research.

Finally, no RCM added value maps are provided for the EM-WAM teleconnection. First, because the contrasting leading SST modes obtained in observations and some model simulations are associated with different teleconnection mechanisms (Figs. 11-13). Thus, it makes no sense to compare precipitation anomalies influenced by ENSO (e.g., observations) with those which are not (e.g., MPI; cf. Table S3). Second, because the underlying dynamics are so mixed between EM and ENSO that assessing causality is not possible at this point.

## 5. Conclusions and discussion

In this article, the ability of GCMs and RCMs to represent the West African Monsoon rainfall regime and its most prominent interannual teleconnections is analyzed. For this purpose, 8 General Circulation Model (GCM) historical simulations from CMIP5 and 11 Regional Climate Model (RCM) simulations from CORDEX-Africa are considered (Giorgi et al. 2009; Nikulin et al. 2012; Taylor et al. 2012). To account for uncertainties in RCM performance and lateral boundary conditions, different GCM/RCM experiment blocks are analyzed (cf. Table 1). First, a set of 8 GCM simulations (a.k.a., GCMs) regionalized on SMHI-RCA4 (GCMs-RCA4). Second, a simulation of MPI-ESM-LR (a.k.a. MPI) driving 4 different RCMs (CCLM4-8, REMO, RCA4 and CRCM5; a.k.a. MPI-RCMs). Observational datasets are also utilized to characterize the West African Monsoon (WAM)/Sea Surface Temperature (SST) teleconnections. The period chosen for the analysis is the peak season of the WAM, July-August-September (JAS), along 26 consecutive years (1979-2004).

Firstly, seasonal biases of mean and standard deviation precipitation fields are characterized (Figs. 1-2). In accordance with previous studies, GCMs and regional simulations have trouble to simulate the northward migration of the Intertropical Convergence Zone (Nikulin et al. 2012; Dosio et al. 2015). This is caused by Atlantic equatorial SST biases present in GCMs (Fig. S1d). In addition, bias improvement in regional simulations is highly sensitive to GCM performance. For instance, MPI-RCM simulations are not able to improve precipitation biases over large areas of West Africa. This could be related to the relative good performance of MPI itself (Figs. 1f and 2) and the fact that convective precipitation is parameterized both in GCMs and RCMs. The opposite is found for GCM and GCM-RCA4 simulations (Figs. 1e and 2). These results may be different in convection-permitting or cloud-resolving simulations (Randall et al. 2003; Prein et al. 2015).

Secondly, the ability of GCMs and RCMs to simulate the most prominent interannual teleconnection patterns of the WAM is assessed. These are, following Rodríguez-Fonseca et al. (2011; 2015): (i) El Niño-Southern Oscillation (ENSO); (ii) the eastern Mediterranean (MED); and (iii) the Atlantic Equatorial Mode (EM). With this aim, a Maximum Covariance Analysis (MCA) is performed using as predictor JAS SST anomalies from: (i) the Equatorial Pacific [110ºE-80ºW, 20ºS-20ºN]; (ii) the Mediterranean [0º-40ºE, 30ºN-45ºN]; and (iii) the equatorial Atlantic [60ºW-20ºE, 20ºS-5ºN], respectively. West African precipitation anomalies over the area [20ºW-30ºE, 0º-20ºN] are considered in all analyses as predictand. For the calculations, the SST based Statistical Seasonal

Forecast model (S4CAST v.2.0), recently developed by Suárez-Moreno and Rodríguez-Fonseca (2015), is utilized. A summary of GCM/RCM performance in representing the most prominent WAM teleconnections is provided below:

1. **ENSO-WAM teleconnection:** In observations, during a positive ENSO event, the anomalous large-scale circulation induces an intensification of air subsidence over West Africa (Fig. 3; the opposite holds for ENSO-; Joly and Voldoire 2009). Thus, precipitation is reduced all along the corridor 0°-20°N (Fig. 3b). The ENSO signal on air subsidence/precipitation is essentially captured by GCMs (Fig. 4 – left column). However, compared to observations, the simulated anomalies appear shifted to the south due to GCM SST biases, and the influence on Sahelian precipitation is weak. Regional simulations (GCM-RCA4 and MPI-RCMs) reveal that the ENSO impact near the equatorial Atlantic is well replicated, even improved over local areas, compared with GCMs. However, over most of West Africa [20°W-30°E, 5°N-20°N], the ENSO influence on air subsidence and precipitation is clearly reduced after regionalization (cf. Figs. 4-5). Hence, RCMs do not appear to be a skillful tool to improve GCM simulations of the ENSO-WAM teleconnection (Fig. 6).

2. **MED-WAM teleconnection:** Sahelian low-level humidity and precipitation are substantially increased when the Mediterranean waters are significantly warmer (Fontaine et al. 2010). As shown in observations, a warmer Mediterranean is associated with negative SLP anomalies over Northern Africa (Saharan Heat Low) and a strong North to South SLP gradient over the Sahel. These conditions promote moisture advection from the Mediterranean (northern branch) and the Gulf of Guinea (southern branch) towards the Sahel (Fig. 7). GCMs have trouble to simulate the strength and location of the Saharan Heat Low. As a consequence, the moisture advection branches impacting the Sahel in observations (Fig. 7c) are misrepresented in GCMs (Figs. 8e and 9e). Particularly, the moisture advection branch from the Gulf of Guinea is the worst simulated. After regionalization, the pressure gradient between the Gulf of Guinea and Sahara is better characterized. As a result, the simulation of southwesterly humidity inflow and precipitation over the Sahel is improved (Figs. 8f and 9f). In this case, RCMs appear to be a skillful tool to improve GCM simulations of the MED-WAM teleconnection (Fig. 10).

3. **EM-WAM teleconnection:** An isolated EM+ SST pattern is known to produce a dipole of precipitation anomalies over West Africa, with positive values in the Gulf of Guinea and negative over the Sahel (Losada et al. 2010). However, due to the period chosen in this study (1979-2004), simultaneous EM+ and ENSO- SST anomalies arise as the leading homogeneous MCA mode from observations (Rodríguez-Fonseca et al. 2009). Due to the combined dynamics of both SST forcings, the heterogeneous precipitation map is no longer a precipitation dipole, but a uniform mode of positive rainfall anomalies extending from the Gulf of Guinea towards the Western Sahel (Fig. 11b; Losada et al. 2010). Very similar SST modes are obtained when MCA is performed on GCMs and GCMs-RCA4 (Figs. 12-13). However, as explained above, GCMs tend to attenuate the ENSO influence on Sahelian precipitation compared to observations, and the signal gets practically removed after regionalization. Therefore, the homogeneous rainfall pattern from observations (Fig. 11b) returns a much more marked dipolar shape as far as GCM (Fig. 12c) and GCM-RCA4 (Fig. 12d) simulations are utilized. Unfortunately, for MPI and MPI-RCMs, the leading co-variability SST modes no longer present statistically significant ENSO- anomalies accompanying EM- (Figs. 13 and S14). As a result, the leading precipitation modes resemble a dipole (Figs. 13c-d) which, by construction, is not logical to compare with observations. Due to the mixed ENSO and EM dynamics in these experiments, no robust conclusions can be determined.

The novel results in this article will help to select the most appropriate RCM simulations to study WAM teleconnections. This outcome is based on the following findings: (1) It has been shown that the ENSO-WAM teleconnection is depreciated in RCMs due to their inability to propagate the ENSO impact on air subsidence along their domain. This result is in line with Boulard et al. (2013), who found over South Africa similar problems with the ENSO teleconnection and attributed these deficiencies to the lateral atmospheric forcing. In this context, the use of spectral nudging to impose the large-scale atmospheric variability within the regional domain might potentially help to improve RCM simulations of the ENSO-WAM teleconnection. Future analyses on RCM architecture, lateral nesting (e.g., Davies-type) and performance should keep the focus on this issue. Since our results also apply for the so-called "ENSO-WAM kelvin wave teleconnection mechanism", forthcoming research on this topic might also consider additional mechanisms proposed in the literature (Joly and Voldoire 2009); (2) The influence of the Mediterranean on the WAM is much better reproduced by RCMs compared to GCMs. Particularly, RCMs improve the representation of the large-scale pressure gradient between the Gulf of Guinea and Sahara, and moisture advection over the Sahel. In this case, a great part of the Mediterranean Sea is included inside the CORDEX-Africa domain (Fig. S1a) and the SST anomalies provided by the forcing GCM are prescribed along the RCM surface. Thus, the adjacent Mediterranean SST forcing appears to notably improve RCM performance. Although for a different region, these results are also in line with Boulard et al. (2013), who found that regional SST forcing over adjacent oceans favored realistic rainfall anomalies over South Africa. Future analyses on this topic should consider the MENA-CORDEX domain, which includes northern Africa, southern

Europe and the whole Arabian Peninsula (Bucchignani et al. 2015). In this context, several studies have just pointed out the importance of a warmer Mediterranean on the recent recovery from the Sahelian drought (Evan et al. 2015; Park et al. 2016); (3) Regarding the EM-WAM teleconnection, the observed mixed EM and ENSO dynamics and the existence of lateral-atmospheric and surface-prescribed SST forcings in the RCM simulations preclude to infer any robust conclusion. The ENSO imprint on this teleconnection could potentially be removed through statistical methods. However, due to the short period considered, the event to event differences and the non-linear interactions between ENSO and other processes (e.g., volcanic eruptions, anthropogenic influence, etc.) these methods result troublesome (Brönnimann 2007). Instead, sensitivity simulations using as predictor an isolated EM mode could be carried out. Much longer observational datasets ought to be considered for this challenge.

Finally, results from GCMs vs. GCMs-RCA4 (lateral boundary conditions) and MPI vs. MPI-RCMs (RCM performance) provide very similar outcomes for the ENSO-WAM and MED-WAM teleconnections. These results are consistent with the main conclusion of this study: the ability of RCMs to represent WAM teleconnections appears to be highly sensitive to the regional domain boundaries and the way the external forcings are prescribed (lateral-atmospheric vs. surface-SST). For a more comprehensive assessment on this topic, additional SST forcing patterns affecting the WAM could be considered in the future (e.g., the Indian ocean and the Tropical North Atlantic; Lu and Delworth 2005; Chung and Ramathan 2006).

*Acknowledgments*

We thank the *National Centers for Environmental Prediction* (NCEP)/*National Center for Atmospheric Research* (NCAR) and the *National Aeronautics and Space Administration* (NASA) for the GPCP and MERRA datasets, respectively. We also thank the *European Centre for Medium-Range Weather Forecasts* and the *Met Office Hadley Centre* for the ERA-Interim and HadISST databases. We are indebted to the *Coupled Model Inter-comparison Project Phase 5* (CMIP5), *Coordinated Regional Climate Downscaling Experiment* (CORDEX) and involved institutions for providing the GCM/RCM simulations used in this study. We also thank the *Earth System Grid Federation* (ESGF) for making these simulations available. This study has been supported by the European Commission's research project PREFACE (EU/FP7 2007-2013; ref. 603521). Iñigo Gómara is also supported by the Spanish Ministry of Economy and Competitiveness ("*Juan de la Cierva-Formación*" contract; FJCI-2015-23874). Finally, we would like to thank the two anonymous reviewers, whose pertinent comments and suggestions have contributed to improve this manuscript.

**Figure Captions:**

**Fig. 1:** (a) Mean (shadings; mm day$^{-1}$) and standard deviation (contours; mm day$^{-1}$) of GPCP JAS seasonal precipitation (1979-2004). Rectangles denote West-Africa North (WA-N), West-Africa South (WA-S) and the whole West Africa (WA) regions. (b) Ensemble biases of GCMs vs. GPCP (1979-2004). Differences in mean/standard deviation are given in colors/contours (mm day$^{-1}$). +/x symbols denote areas where all models depict the same sign in mean/standard deviation biases. (c) Same as (b) but for the ensemble of GCM-RCA4 simulations. (d) Same as (b) but for the ensemble of MPI-RCM simulations. (e) Ensemble added value in mean/standard deviation of GCM-RCA4 simulations in colors/contours (mm$^2$ day$^{-2}$). 75% of members giving the same added value sign in mean/standard deviation are marked with +/x. (f) Same as (e) but for MPI-RCMs.

**Fig. 2:** (a) Biases in mean precipitation values averaged over West Africa – North [10°W-10°E, 9°N-15°N] with respect to observations (GPCP – horizontal offset; mm day$^{-1}$). Blue bars correspond to GCM biases. Yellow bars correspond to GCM-RCA4 biases. Red bars correspond to MPI-RCM biases. The horizontal red line provides the climatological value from MERRA. (b) Same as (a) but for standard deviation biases over West-Africa – North (mm day$^{-1}$). (c)-(d) Same as (a)-(b) but for the West Africa – South region [10°W-10°E, 5°N-9°N]. (e)-(f) Same as (a)-(b) but for the whole West Africa region [20°W-30°E, 0°N-20°N].

**Fig. 3:** (a) JAS SST anomalies (HadISST) regressed on expansion coefficient U in colors (homogeneous map: K std$^{-1}$). 90% confidence interval in stippling (Monte Carlo test: 1000 random iterations). Differences in velocity potential anomalies at 200 hPa with respect to those at 850 hPa (KHI200/850; ERA-Interim) regressed on U in contours (every 0.8 10$^6$ m$^2$ s$^{-1}$). (b) JAS precipitation anomalies (GPCP) regressed on U in colors (heterogeneous map; mm day$^{-1}$ std$^{-1}$). 90% confidence interval in stippling. (c) Differences in wind divergence anomalies at 200 hPa with respect to those at 850 hPa (DIV200/850; ERA-Interim) regressed on U in colors (10$^{-6}$ s$^{-1}$ std$^{-1}$). 90% confidence interval in stippling. Differences in velocity potential anomalies at 200 hPa with respect to those at 850 hPa (KHI200/850; every 0.4 10$^6$ m$^2$ s$^{-1}$) regressed on U in contours. Expansion coefficient U is obtained from a Maximum Covariance Analysis. Squared covariance fraction provided in %. Predictor: SST [110°E-80°W, 20°S-20°N; dashed rectangle in (a)]. Predictand: PCP [20°W-30°E, 0°-20°N; dashed rectangle in (b)]. Period: 1979-2004.

**Fig. 4:** (a)-(c)-(e) Same as Fig. 3 but for the ensemble mean of GCMs (stippling - 7/8 models with same sign on regression in SST, rainfall and DIV200/850). (b)-(d)-(f) Same as (a)-(c)-(e) but for the ensemble mean of GCMs-RCA4.

**Fig. 5:** (a)-(c)-(e) Same as Fig. 3 but for the MPI simulation. (b)-(d)-(f) Same as (a)-(c)-(e) but for the ensemble mean of MPI-RCMs. 4/4 models with same sign on regression: shadings in (b), stippling in (d)-(f).

**Fig 6:** (a) Ensemble added value (AV) map (colors, in $mm^2$ $day^{-2}$ $std^{-2}$) of the ENSO-WAM teleconnection for precipitation, calculated as the average of AV values from each individual model. Positive values: areas where GCMs-RCA4 improve GCM simulations of the ENSO-WAM teleconnection (considering observations as basis). Red/Blue contours: number of simulations of each experiment with positive AVs over each grid point. (b) Same as (a) but for MPI-RCM simulations.

**Fig 7:** Same as Fig. 3 but using as predictor Mediterranean SST anomalies [0º-40ºE, 30ºN-45ºN]. No KHI200/850 regression anomalies are plotted. Regression maps on (c) are based on anomalous specific humidity at 850 hPa (SHUM850 - colors, in $Kg$ $Kg^{-1}$ $std^{-1}$; 90% confidence level in green circles), sea level pressure (SLP - contours, $hPa$ $std^{-1}$) and moisture flux at 850 hPa (MF850 - arrows, $Kg$ $Kg^{-1}$ $m$ $s^{-1}$ $std^{-1}$).

**Fig. 8:** (a)-(c)-(e) Same as Fig. 7 but for the ensemble mean of GCMs (stippling/circles - 7/8 models with same sign on regression in SST, precipitation and SHUM850). (b)-(d)-(f) Same as (a)-(c)-(e) but for the ensemble mean of GCMs-RCA4.

**Fig. 9:** (a)-(c)-(e) Same as Fig.7 but for the MPI simulation. (b)-(d)-(f) Same as (a)-(c)-(e) but for the ensemble mean of MPI-RCMs. 4/4 models with same sign on regression: shadings in (b), stippling/circles in (d)-(f)

**Fig. 10:** Same as Fig. 6 but for the Mediterranean-WAM teleconnection.

**Fig 11:** Same as Fig. 3 but using as predictor Atlantic SST anomalies [60ºW-20ºE, 20ºS-5ºN]. Regression maps on (c) are based on anomalous specific humidity at 850 hPa (SHUM850 - contours, in $Kg$ $Kg^{-1}$ $std^{-1}$), moisture

flux at 850 hPa (MF850 - vectors, Kg Kg$^{-1}$ m s$^{-1}$ std$^{-1}$) and wind divergence difference between 200/850 hPa (DIV200/850 - colors, 10$^{-6}$ s$^{-1}$ std$^{-1}$; 90% confidence interval in white circles).

**Fig. 12:** (a)-(c)-(e) Same as Fig. 11 but for the ensemble mean of GCMs (stippling/circles - 7/8 models with same sign on regression in SST, precipitation and DIV200/850). (b)-(d)-(f) Same as (a)-(c)-(e) but for the ensemble mean of GCMs-RCA4.

**Fig. 13:** (a)-(c)-(e) Same as Fig. 11 but for the MPI simulation. (b)-(d)-(f) Same as (a)-(c)-(e) but for the ensemble mean of MPI-RCMs. 4/4 models with same sign on regression: shadings in (b), stippling/circles in (d)-(f)**.**

**SEASONAL PRECIPITATION BIASES**

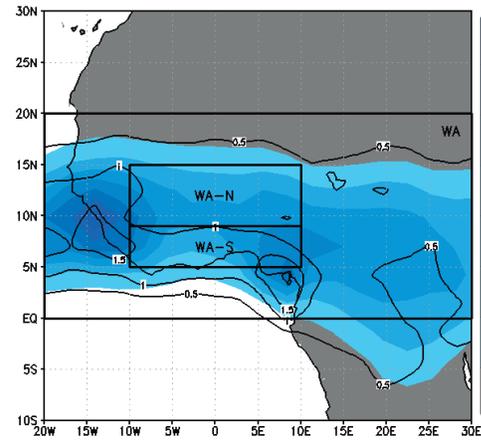
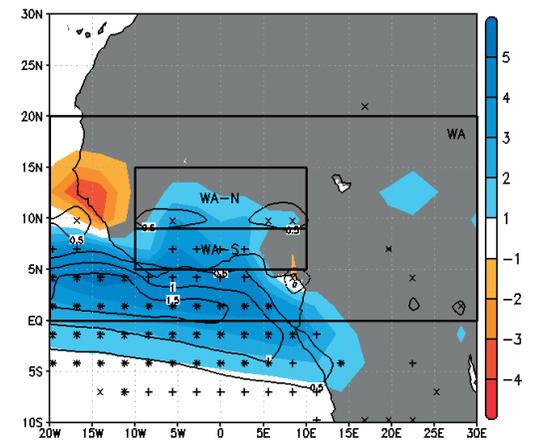
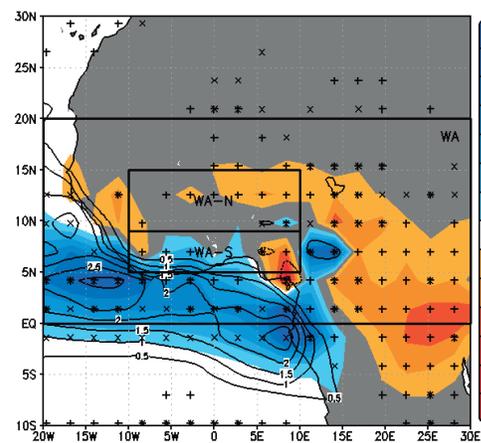
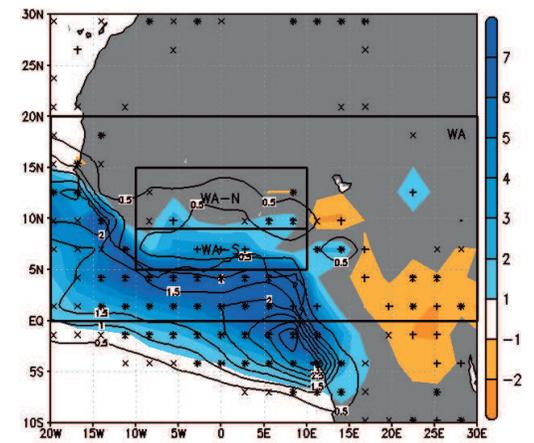
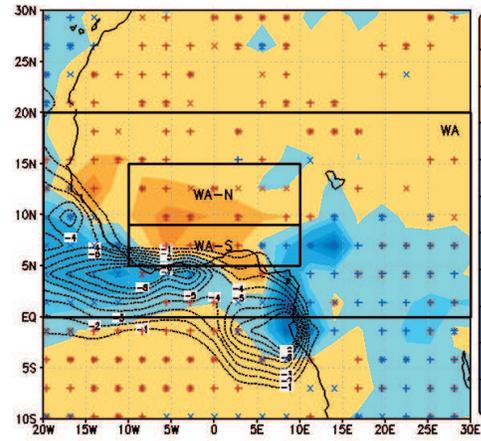
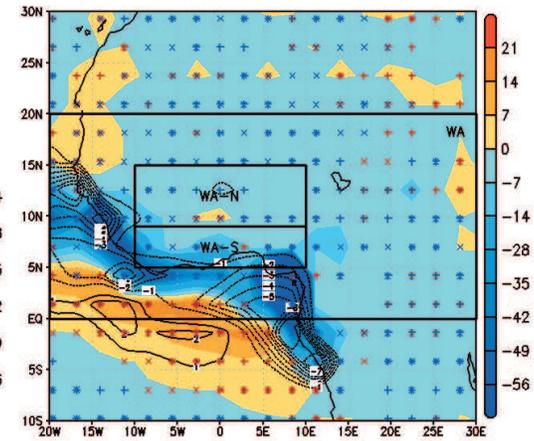

Figure2

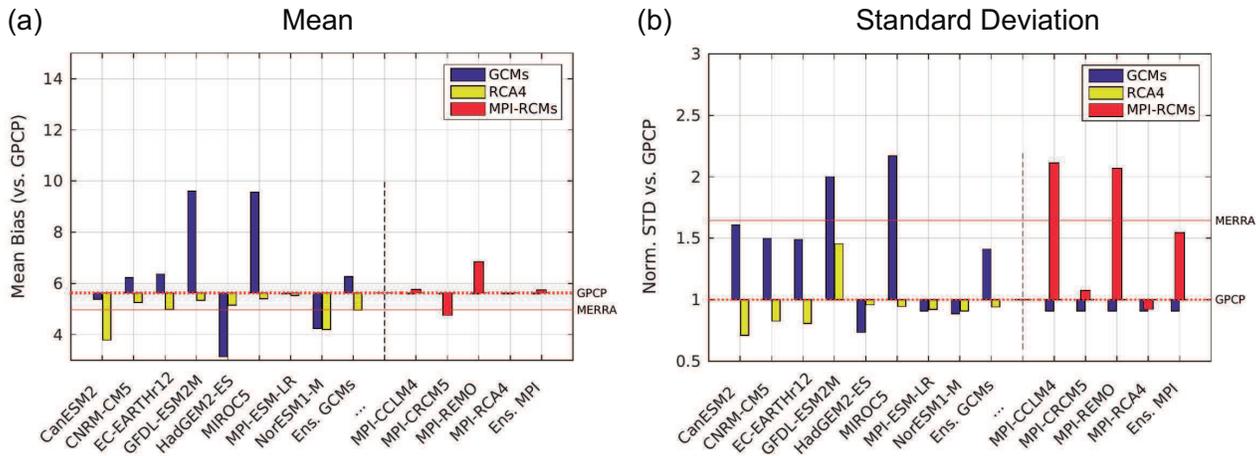

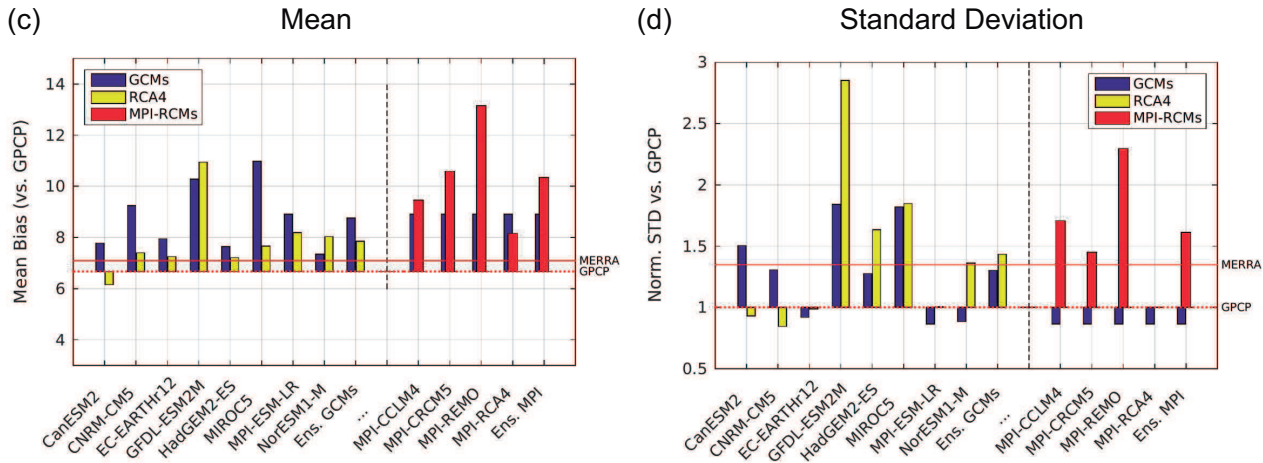

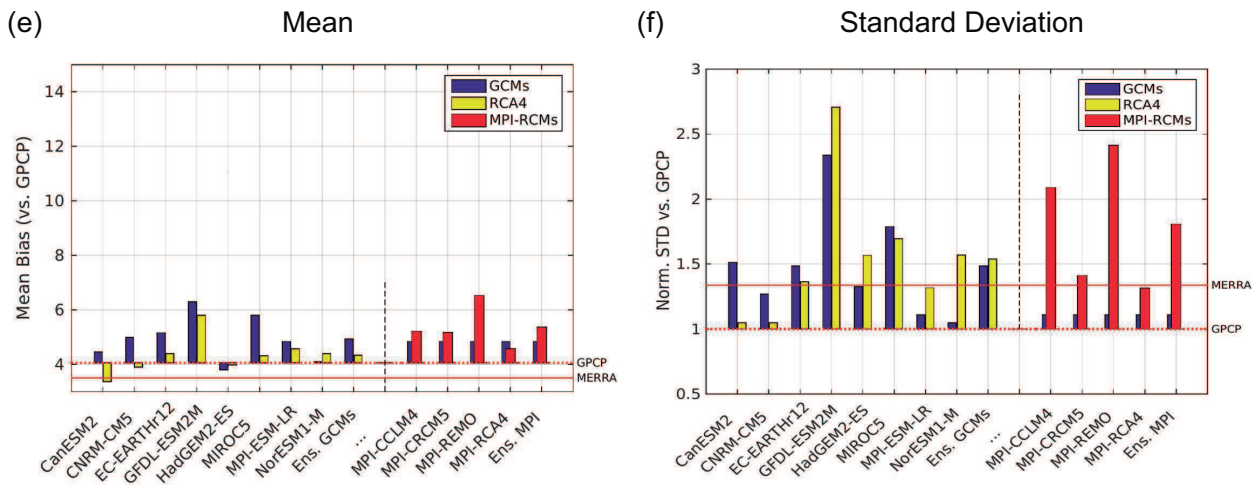

Figure3

## MCA LEADING MODE: OBSERVATIONS

(a) Predictor: HadISST (41%)

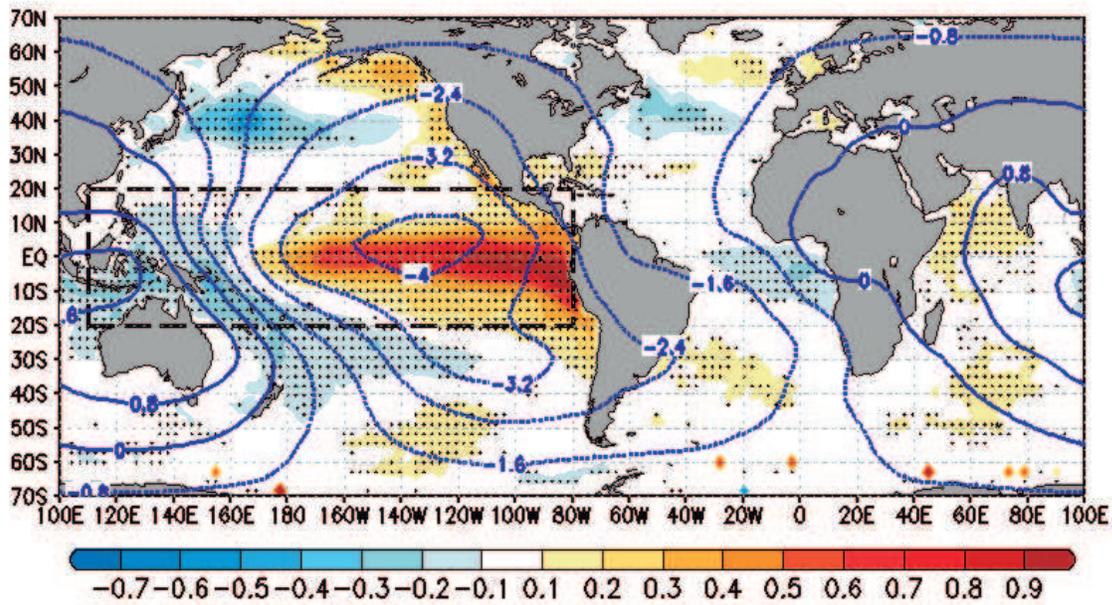

(b) Predictand: GPCP

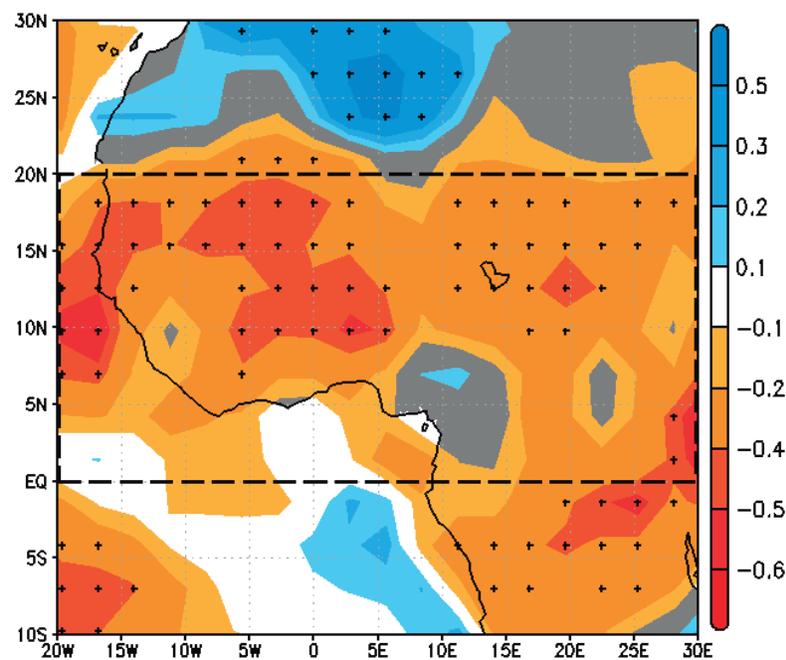

(c) U Regression Maps (ERA-Int.)

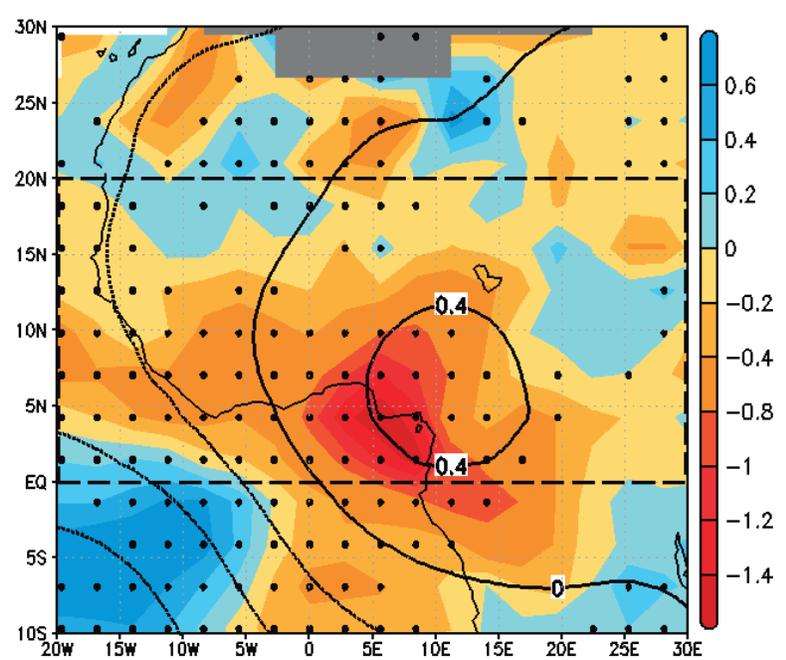

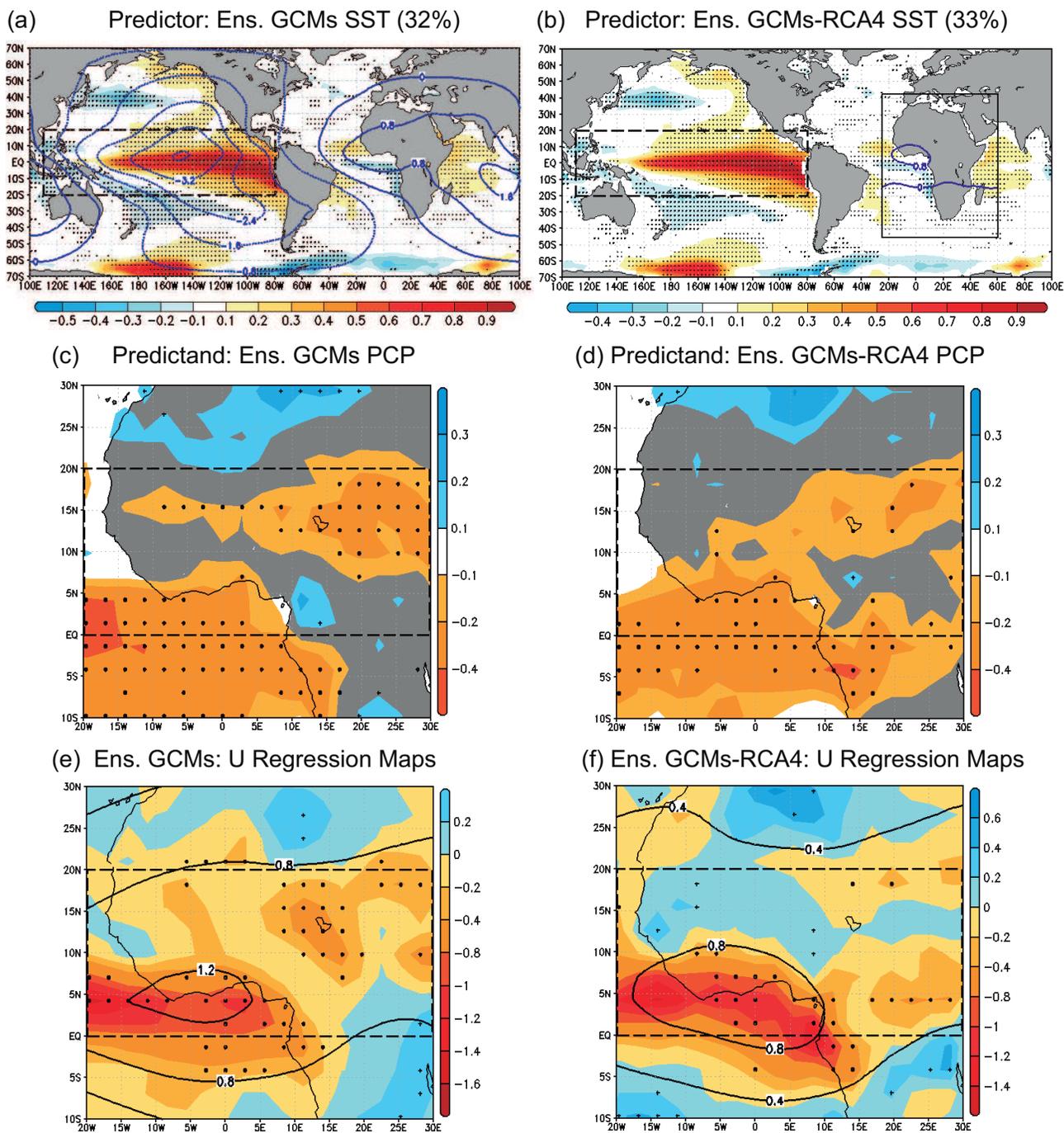

Figure 4. MCA LEADING MODES: GCMs & GCMs-RCA4

Figure5
**MCA LEADING MODES: MPI & MPI-RCMs**

(a) Predictor: MPI SST (33%)  (b) Predictor: Ens. MPI-RCMs SST (31%)

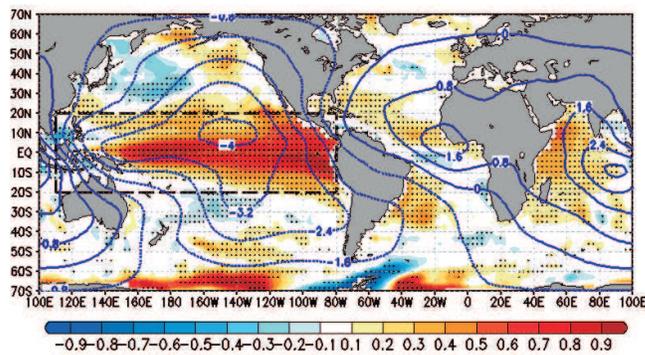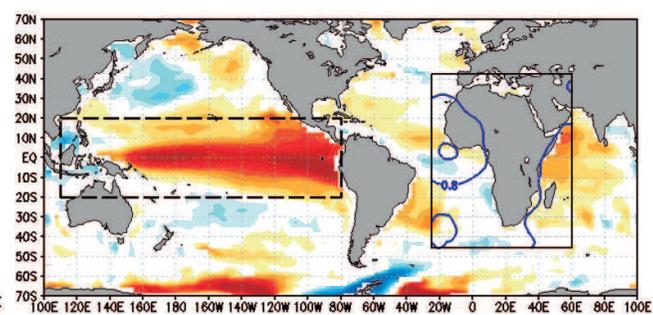

(c) Predictand: MPI PCP  (d) Predictand: Ens. MPI-RCMs PCP

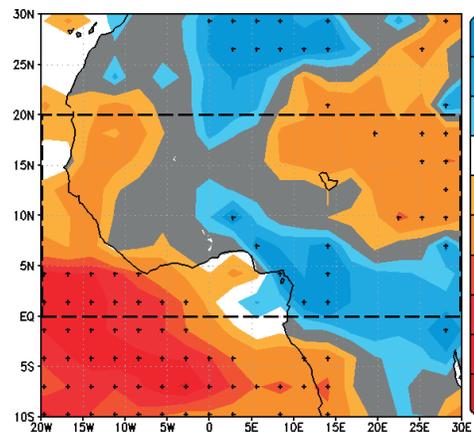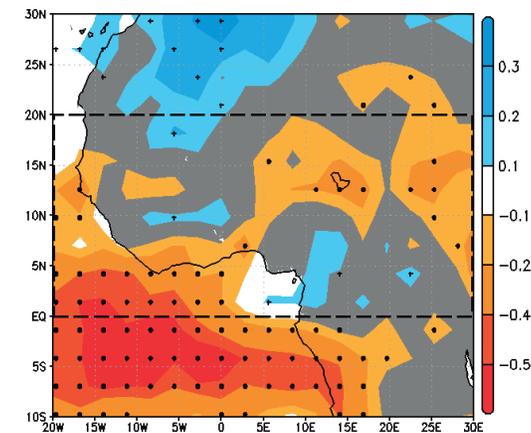

(e) MPI: U Regression Maps  (f) Ens. MPI-RCMs: U Regression Maps

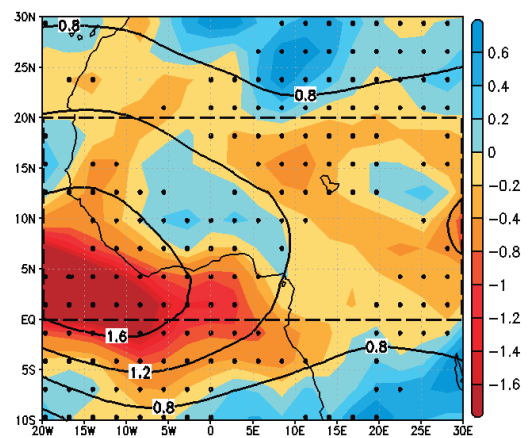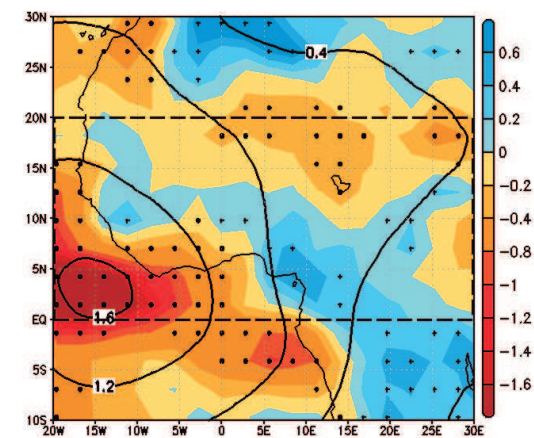

Figure6 **ENSO-WAM TELECONNECTION: ENSEMBLE ADDED VALUES**

(a) GCMs-RCA4 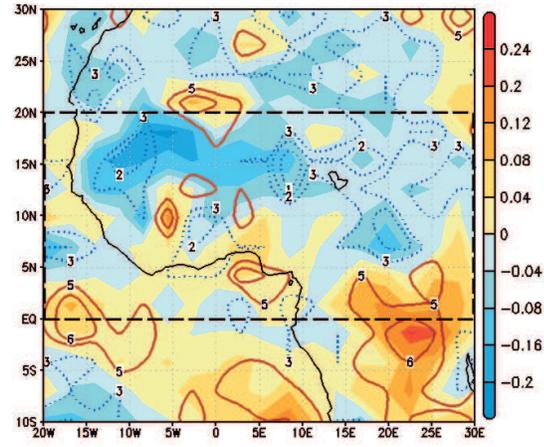

(b) MPI-RCMs 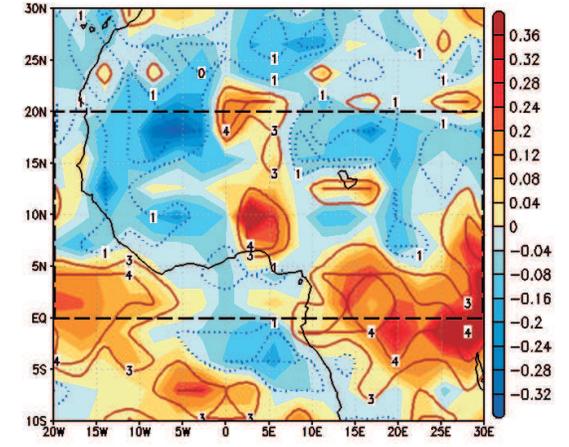

Figure7

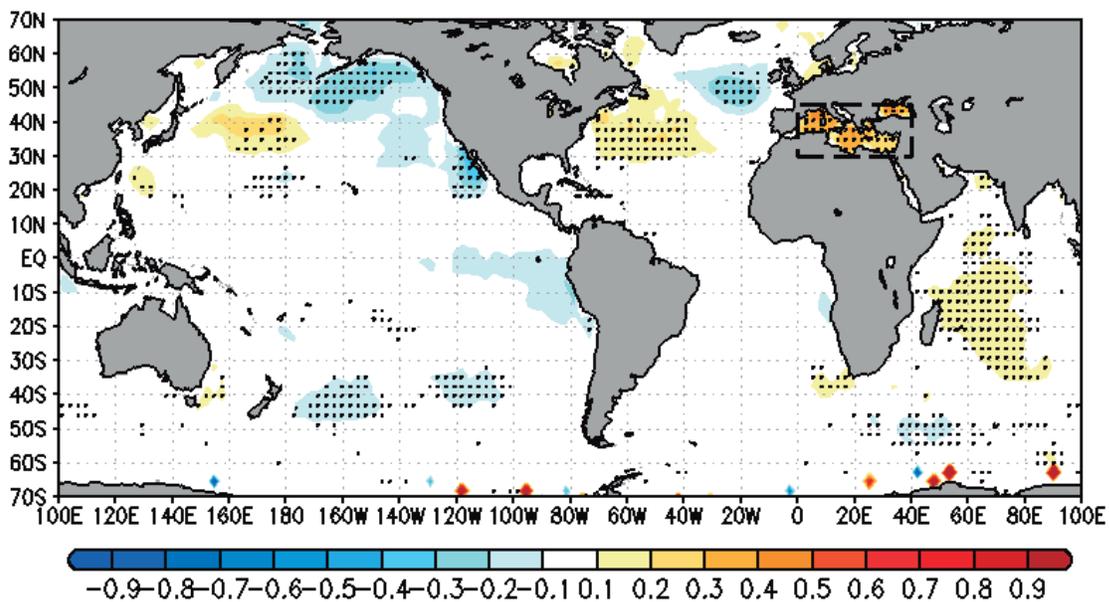

**MCA LEADING MODE: OBSERVATIONS**
(a) Predictor: HadISST (40%)

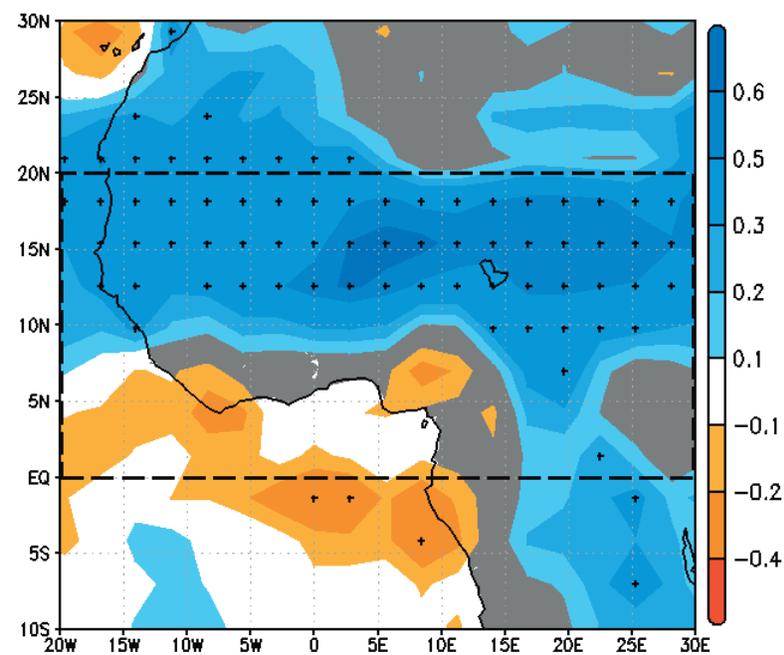

(b) Predictand: GPCP

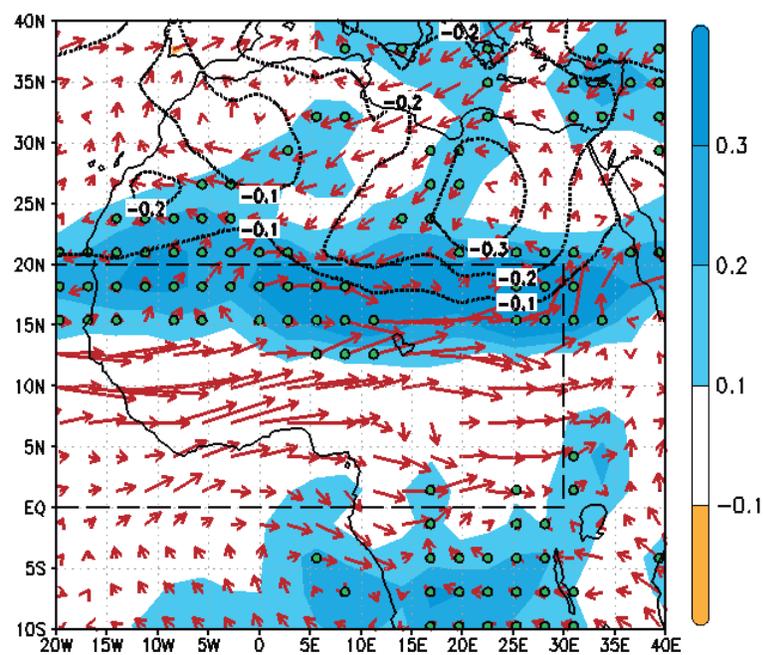

(c) U Regression Maps (ERA-Int.)

Figure8

**MCA LEADING MODES: GCMs & GCMs-RCA4**

(a) Predictor: Ens. GCMs SST (32%)  (b) Predictor: Ens. GCMs-RCA4 SST (33%)

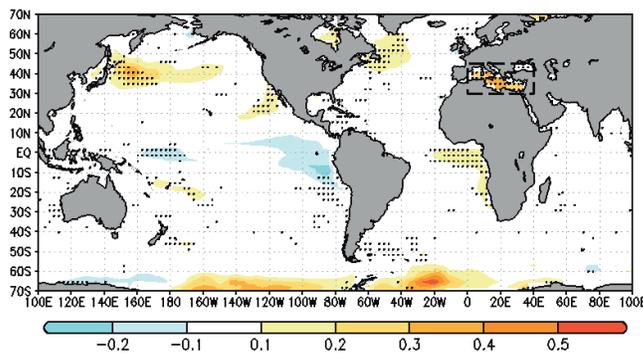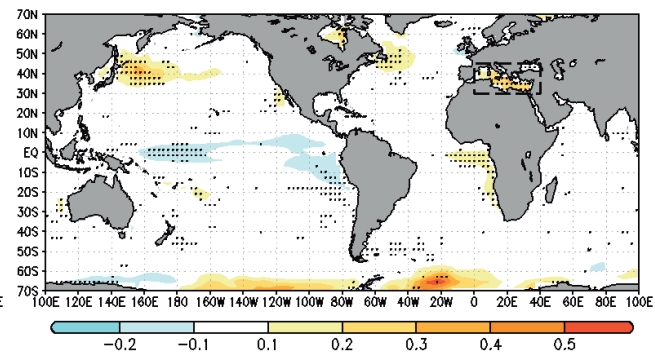

(c) Predictand: Ens. GCMs PCP  (d) Predictand: Ens. GCMs-RCA4 PCP

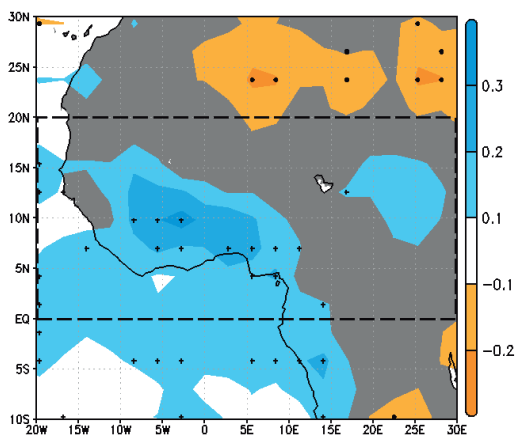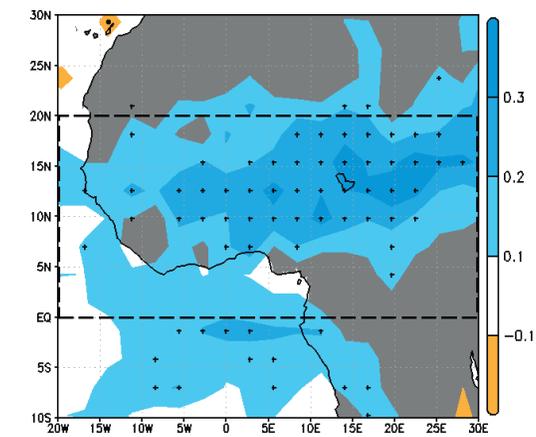

(e) Ens. GCMs: U Regression Maps  (f) Ens. GCMs-RCA4: U Regression Maps

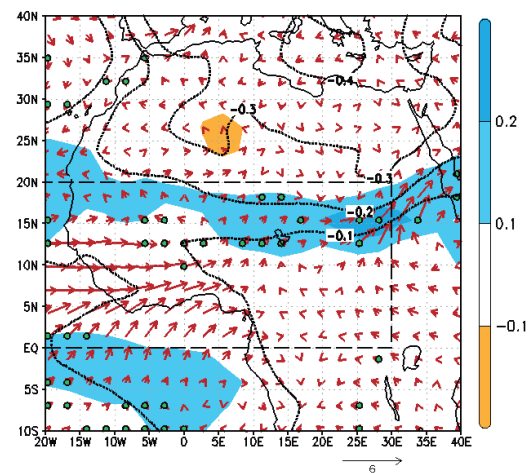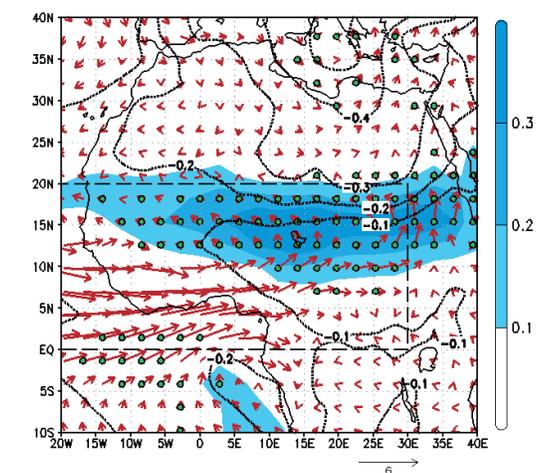

Figure9

**MCA LEADING MODES: MPI & MPI-RCMs**

(a) Predictor: MPI SST (29%)  (b) Predictor: Ens. MPI-RCMs SST (26%)

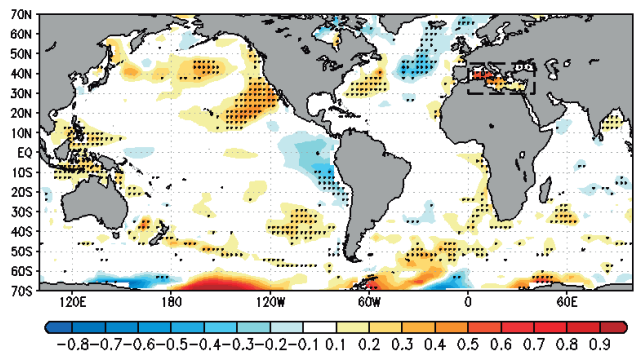
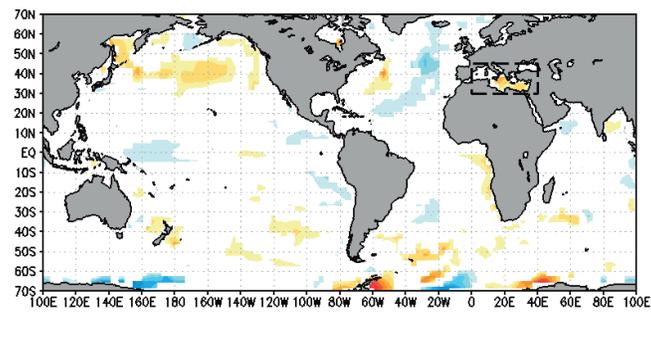

(c) Predictand: MPI PCP  (d) Predictand: Ens. MPI-RCMs PCP

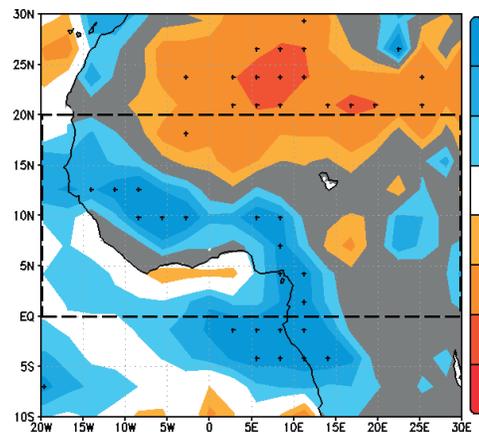
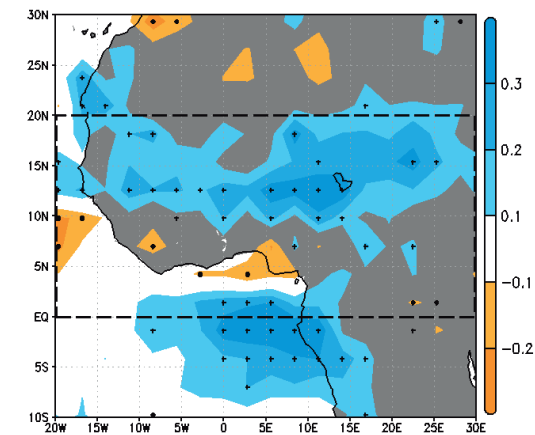

(e) MPI: U Regression Maps  (f) Ens. MPI-RCMs: U Regression Maps

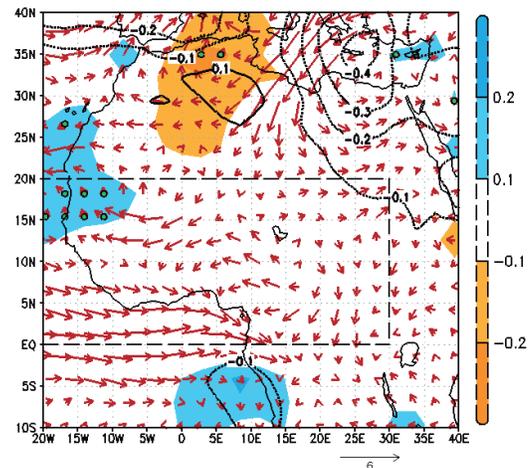
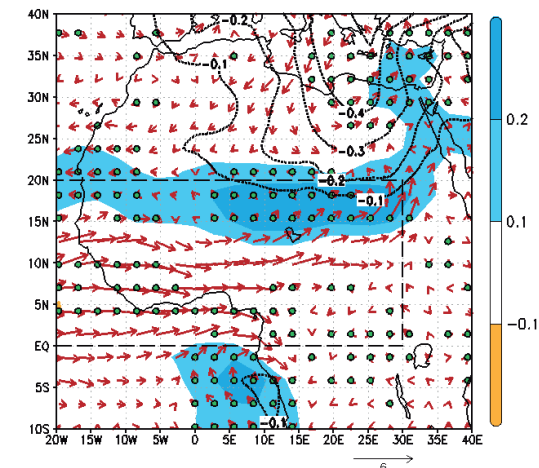

**MED-WAM TELECONNECTION: ENSEMBLE ADDED VALUES**

(a) GCMs-RCA4  (b) MPI-RCMs

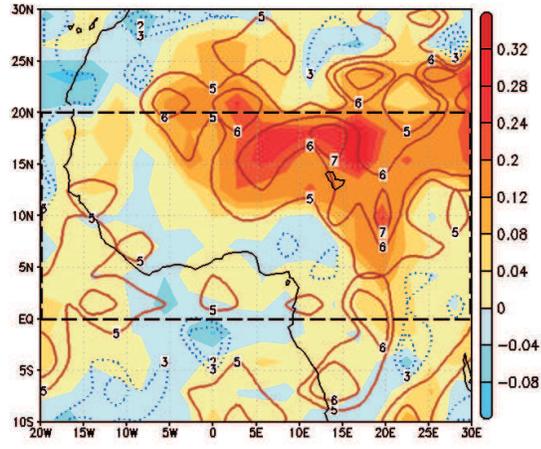
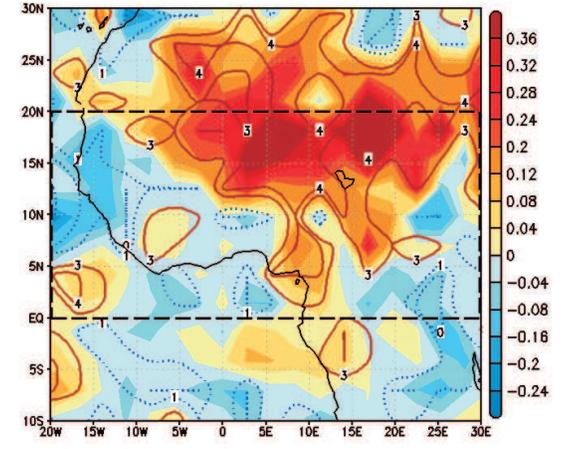

Figure11

## MCA LEADING MODE: OBSERVATIONS

(a) Predictor: HadISST (41%)

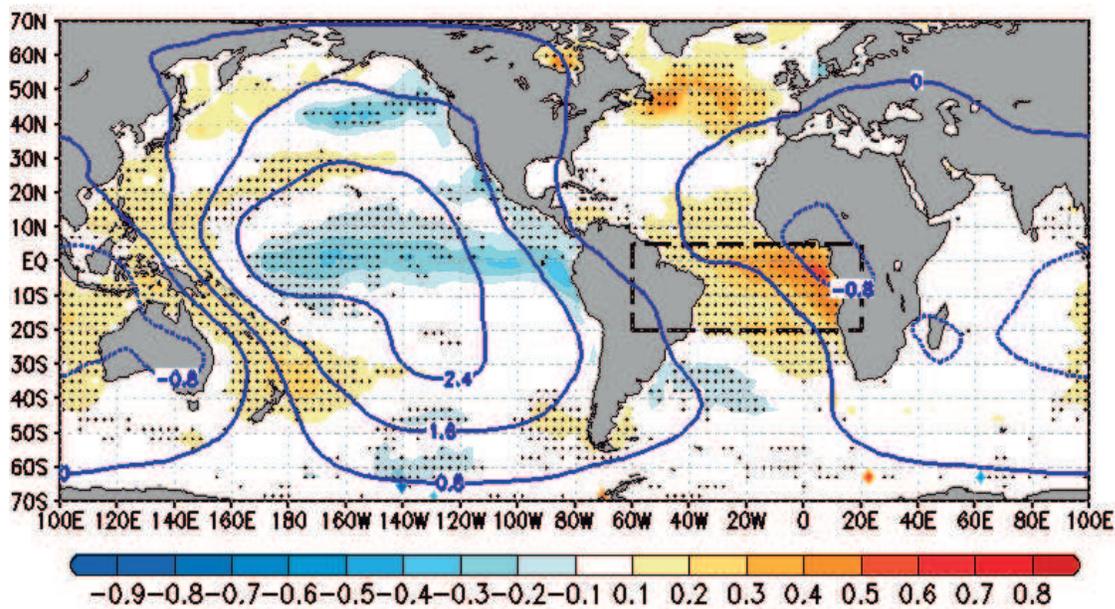

(b) Predictand: GPCP

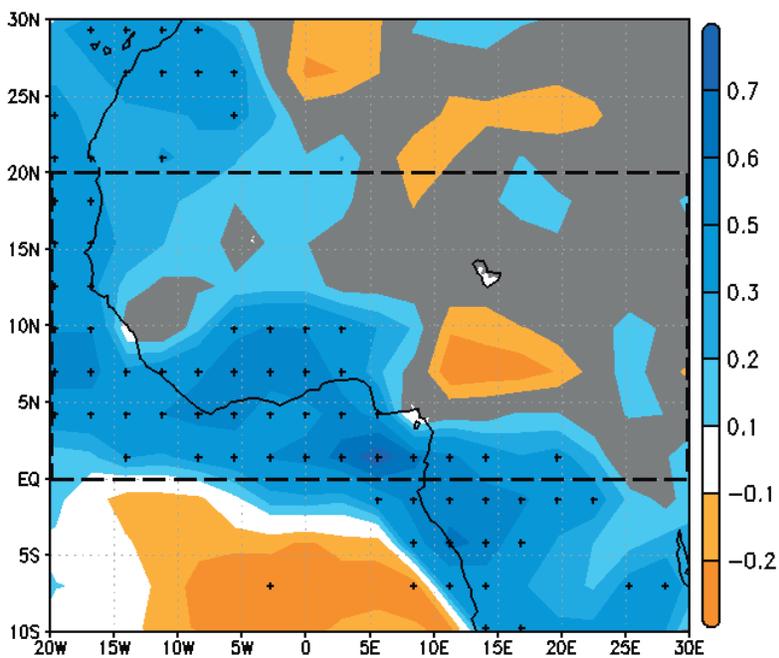

(c) U Regression Maps (ERA-Int.)

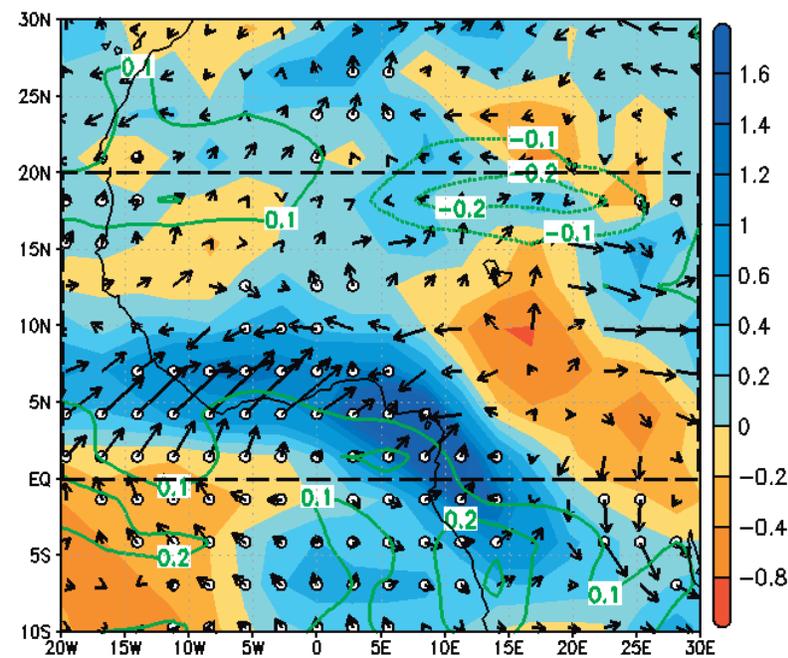

Figure12

**MCA LEADING MODES: GCMs & GCMs-RCA4**

(a) Predictor: Ens. GCMs SST (36%)  (b) Predictor: Ens. GCMs-RCA4 SST (34%)

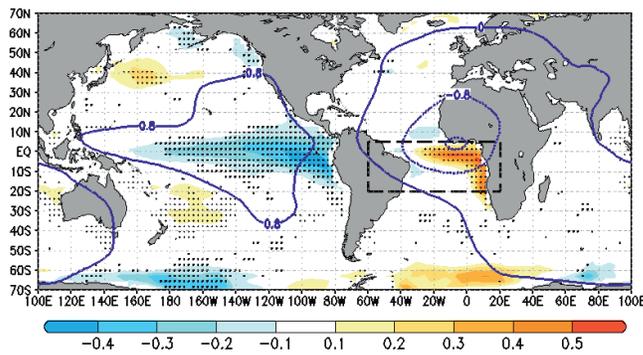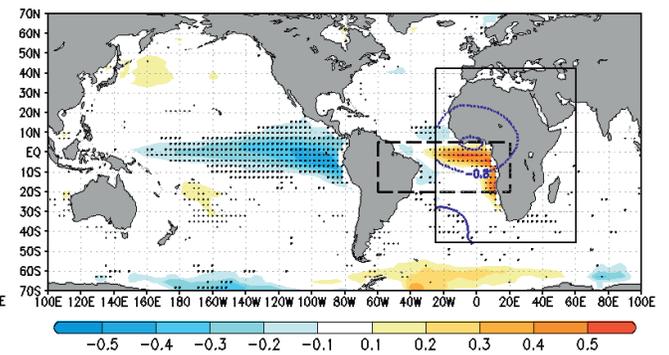

(c) Predictand: Ens. GCMs PCP  (d) Predictand: Ens. GCMs-RCA4 PCP

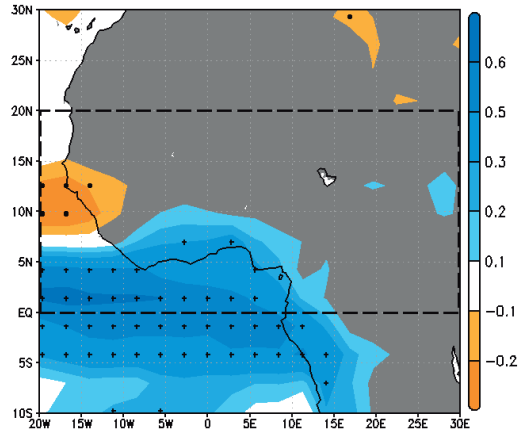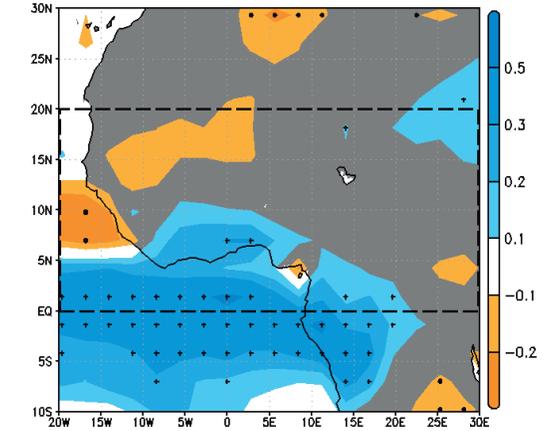

(e) Ens. GCMs: U Regression Maps  (f) Ens. GCMs-RCA4: U Regression Maps

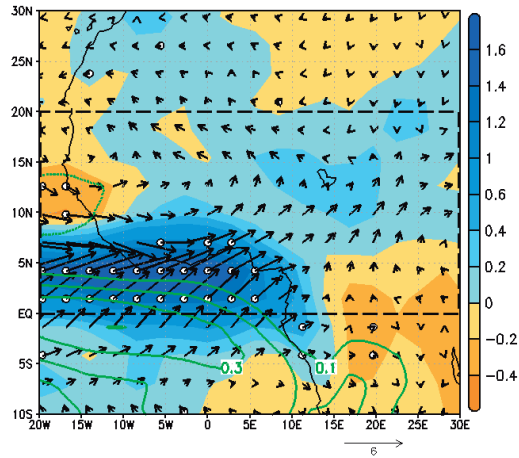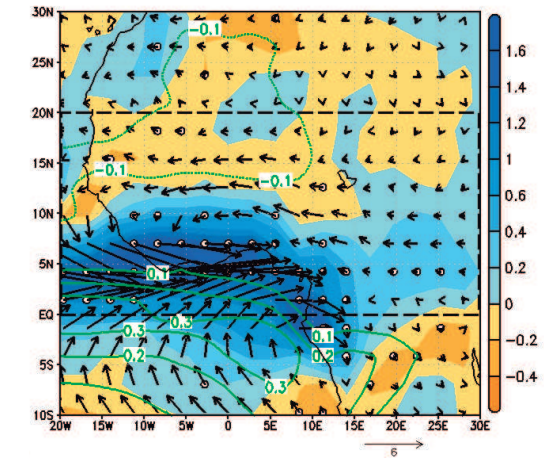

Figure 13

**MCA LEADING MODES: MPI & MPI-RCMs**

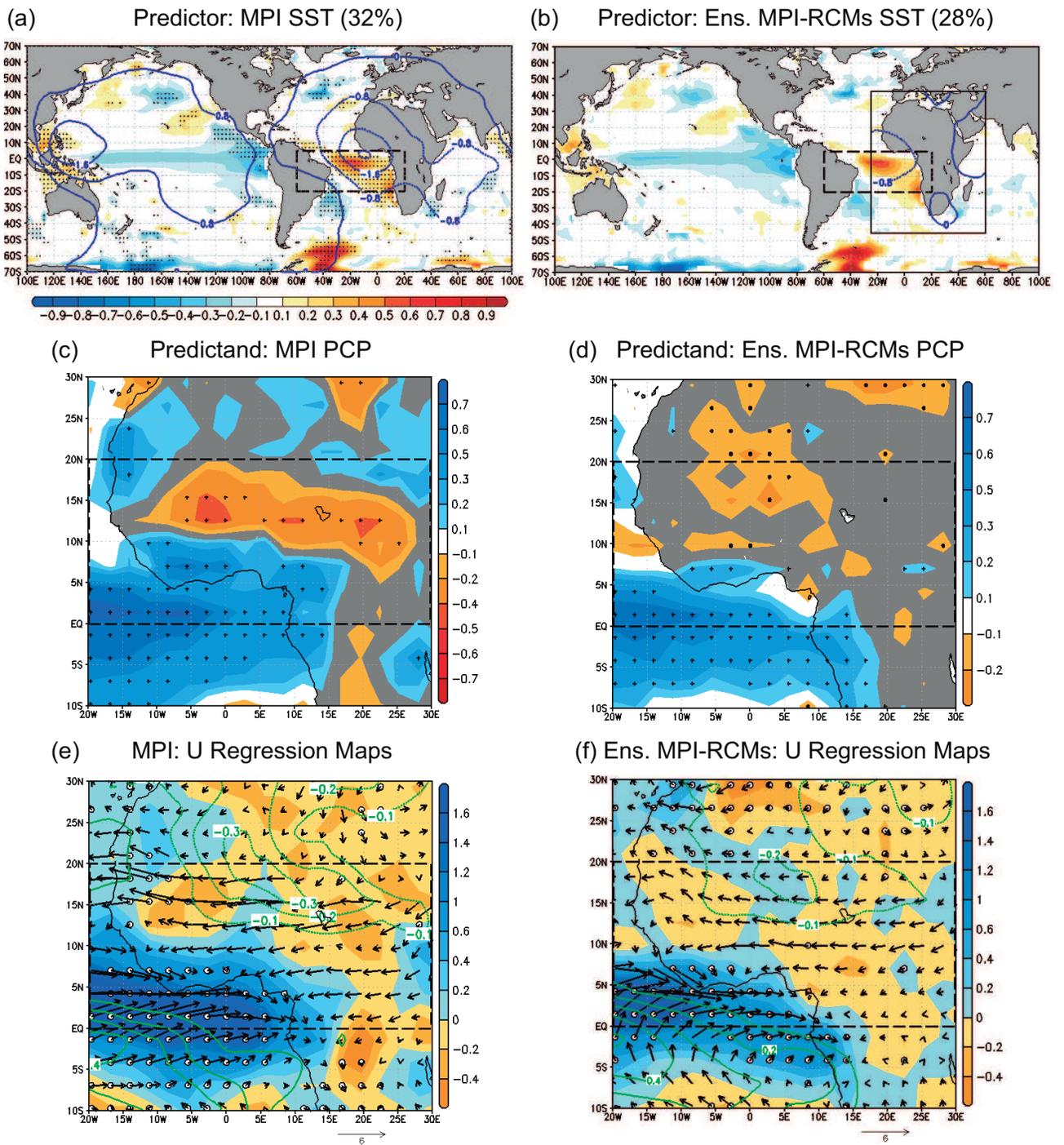

**Tables:**

**Table 1:** CMIP5/CORDEX-Africa historical simulations matrix.

| Historical CMIP5 GCMs / CORDEX-Africa RCMs Matrix | **CanESM2** *CCMA-Canada* | **CNRM-CM5** *CNRM-France* | **EC-EARTH (r12)** *ECMWF-European* | **GFDL-ESM2M** *NOAA GFDL-USA* | **HadGEM2-ES** *Met Office-UK* | **MIROC5** *JAMSTEC-Japan* | **MPI-ESM-LR** *MPI-Germany* | **NorESM1-M** *NCC-Norway* |
|---|---|---|---|---|---|---|---|---|
| **CLMcom-CCLM4-8** *CCLM community-International* | | | | | | | ■ | |
| **CSC-REMO** *MPI-Germany* | | | | | | | ■ | |
| **SMHI-RCA4** *SMHI-Sweden* | ■ | ■ | ■ | ■ | ■ | ■ | ■ | ■ |
| **UQAM-CRCM5** *UQAM-Canada* | | | | | | | ■ | |

CORDEX-Africa RCMs (Giorgi et al. 2009) and CMIP5 GCMs (Taylor et al. 2012) used. Shaded cells indicate selected GCM-RCM combinations. All data were retrieved from the Earth System Grid Federation (ESGF) portals, regridded to the CanESM2 horizontal resolution (coarser one; 2.8º x 2.8º) and selected for the period July-August-September 1979-2004. The Institute ID of each model is provided in bold italics.